\documentclass[journal,onecolumn]{IEEEtran}
\newlength{\doublecolumnwidth}
\usepackage{cite}

\ifCLASSINFOpdf
  \usepackage[pdftex]{graphicx}
\else
  \usepackage[dvips]{graphicx}
\fi
\usepackage[cmex10]{amsmath}
\usepackage{amsfonts}
\usepackage{color}

\def\eq#1{Eq.~\eqref{eq:#1}}
\def\fig#1{Fig.~\ref{fig:#1}}
\def\eq#1{Eq.~\eqref{eq:#1}}

\def\eq#1{Eq.~\eqref{eq:#1}}
\def\fig#1{Fig.~\ref{fig:#1}}
\def\sec#1{Sec.~\ref{sec:#1}}

\newtheorem{definition}{Definition} 
\newtheorem{lemma}{Lemma} 
\newtheorem{theorem}{Theorem} 
\newtheorem{proposition}{Proposition}

\begin{document}
%
\title{Convolutional Polar Codes}
%
%
%

\author{Andrew~James~Ferris, Christoph~Hirche and~David~Poulin
\thanks{While most of this work was completed, A. Ferris was with D\'epartement de Physique, Universit\'e de Sherbrooke,
Sherbrooke, Qu\'ebec J1K 2R1, Canada. He was then with Institut de Ciencies Fotoniques, Parc Mediterrani de la Tecnologia, 08860 Barcelona, Spain as well as Max-Planck-Institut f\"ur Quantenoptik, Hans-Kopfermann-Str. 1, 85748 Garching, Germany, and is currently at Fugro Roames, Qld Australia 4113.}
\thanks{C. Hirche is with the F\'{\i}sica Te\`{o}rica: Informaci\'{o} i Fen\`{o}mens
 Qu\`{a}ntics, 
 Departament de F\'{i}sica, 
Universitat Aut\`{o}noma de Barcelona,   ES-08193
 Bellaterra (Barcelona), Spain.}
\thanks{D. Poulin is with the D\'epartement de Physique \& Institut Quantique, Universit\'e de Sherbrooke,
Sherbrooke, Qu\'ebec J1K 2R1, Canada.}
}

\maketitle

\begin{abstract}
Arikan's Polar codes  \cite{A09a} attracted much attention as the first efficiently decodable and capacity achieving codes. Furthermore, Polar codes exhibit an exponentially decreasing block error probability with an asymptotic error exponent upper bounded by $\beta < \frac{1}{2}$. Since their discovery, many attempts have been made to improve the error exponent and the finite block-length performance, while keeping the bloc-structured kernel. In  \cite{FP14}, two of us introduced a new family of efficiently decodable error-correction codes based on a recently discovered efficiently-contractible tensor network family in quantum many-body physics \cite{EV12b}, called branching MERA. These codes, called branching MERA codes, include Polar codes and also extend them in a non-trivial way by substituting the bloc-structured kernel by a convolutional structure. Here, we  perform an in-depth study of a particular example that can be thought of as a direct extension to Arikan's Polar code, which we  therefore name {\em Convolutional Polar codes}. We prove that these codes polarize and exponentially suppress the channel's error probability, with an asymptotic error exponent $\beta' < \frac{\log_2{3}}{2}$ which is provably better than for Polar codes under successive cancellation decoding. We also perform finite block-size numerical simulations which display improved error-correcting capability with only a minor impact on decoding complexity.
\end{abstract}


%
\IEEEpeerreviewmaketitle

\section{Introduction}
%
%
%
%

The phenomenon of channel polarization, discovered by Arikan \cite{A09a}, can be produced by a controlled-NOT (CNOT) gate. Because the control bit is added to the target bit, it becomes redundantly encoded and thus more robust to noise. On the other hand, the information of the target bit is partially washed away because its value is modified in a way that depends on the value of the possibly unknown control bit. We thus say that the channels have partially polarized into a better and a worse channel. The encoding circuit of a Polar code is obtained by iterating this polarization procedure, and asymptotically produces a perfect polarization, where a fraction of the channels are error-free and the complement are completely randomizing. 

Because of this recursive nature, the encoding circuit takes the geometric form of a spectral transformation where CNOT gates follow a hierarchical arrangement on different length scales (depicted in \fig{circuits} (a)), and much like the (fast) Fourier transform, the linear encoding matrix can be decomposed into a Kronecker product of small matrices.  In this case, the polarization is defined with respect to the successive cancellation decoder, where the marginal probability of input bit $i$ is calculated with the prior knowledge of the bits $1,\dots,i\!-\!1$. 

 In \cite{FP14}, two of the current authors introduced a broad family of codes building on the powerful graphical calculus developed in the field of quantum many-body physics. In this field, the encoding circuit associated to Polar codes are a restricted form of the {\em branching multi-scale entanglement renormalization ansatz} (branching MERA) tensor networks~\cite{EV12b}. More precisely, they correspond to branching MERA networks with half of the tensors being trivial identity gates, resulting in an object that could be called a `branching tree'. The branching MERA code family, defined in \cite{FP14}, contains all the codes obtained from an encoding circuit with the topology of the branching MERA, and includes the Polar code as a special case with many trivial gates. In this Article, we focus on the specific branching MERA code which is the next simplest example obtained by reinserting the missing tensors in the Polar code network. While Polar codes use a sequence of polarization steps, each composed of a product of gates operating on non-intersecting, finite-size blocs, the generalization we consider here goes beyond this scheme by using a convolutional structure at every polarization step. We thus name them {\em Convolutional Polar codes}.

With Polar codes, Arikan was able to give the first constructive example of a provably efficient and capacity-achieving code for symmetric channels \cite{A09a}, generating significant interest and stimulating further work on improvements and generalizations, e.g., \cite{AT09a,STA09a,KU10a,KSU10a,MV11a}.  In particular, the limitations of Polar Codes have been investigated, with the notable result~\cite{KSU10a} that while keeping the decomposability into Kronecker products, the asymptotic block error performance of Polar codes, with an error exponent $\beta = \frac 12 $, is optimal considering underlying matrices with small dimension. In fact, the dimension has to be raised to at least 15 to improve the asymptotic error exponent. Also in~\cite{KSU10a} it was shown that the optimal $16\times16$ matrix achieves an error exponent of $0.51828$. The downside of this approach is that the complexity of encoding and decoding Polar codes with large generating matrices grows with the size of the matrix and are given by $O(lN\log_lN)$ and $O(\frac{2^l}{l}N\log_lN)$ for an $l\times l$ matrix, respectively. Convolutional Polar codes re-open the investigation of optimal behavior by abandoning the bloc-structure-limited polarization in favor of a more general polarization structure.
 
As we will show, Convolutional Polar codes form a natural generalization of Polar codes, and inherit many of their properties including a successive cancellation decoder that  produces a tensor contractible in log-linear time.  While the decoding algorithm is slower by a small, constant numerical factor, we observe a significant improvement in both the channel polarization and the error-correction performance. While an important practical limitations of Polar codes are their finite-size effects \cite{KU10a}, we observe that Convolutional Polar codes display a steeper waterfall region, thus suppressing such finite-size effects.

While compelling numerical evidence provided in \cite{FP14}, and extended here, indicate that Convolutional Polar codes achieve better finite-size performances that Polar codes, formal proofs were not given, including the proposition that Convolutional Polar codes are actually capacity achieving. In the meanwhile investigation of Polar codes made significant progress and many new techniques have been developed. Here we focus on two of them, one being a conceptually simpler proof of polarization~\cite{AT14}, based on the so-called \textit{Mrs. Gerbers Lemma} instead of the Martingale approach used by Arikan. The other one being tools for developing more refined bounds on the speed of polarization and the block error probability, first used in \cite{AT09} and further developed e.g. in \cite{GX15}. We  use these tools to show that Convolutional Polar codes are capacity achieving and  polarize the  channels with an exponentially decreasing block error rate. Furthermore we will see that in the limit of very large block length, the error exponent $\beta' = \frac 12 \log_23\approx 0.79$ is provably better than it is for Polar codes $\beta = \frac 12 $.


\subsection{Paper outline}

This paper is structured as follows. While  branching MERA codes, introduced in \cite{FP14}, form a broad family defined by the topology of their encoding circuit, this Article focuses on a specific example, which we henceforth refer to as ``the Convolutional polar code'', and more specifically all numerical experiments and proofs focus on the open boundary code. In the next Section, we give a brief summary of the tensor network formalism and recast the decoding problem in this language. Then, in \sec{geo}, we detail the encoding circuit for Convolutional Polar codes, comparing it to the Polar code, and describe how to implement the successive-cancellation decoder.  We also address the delicate boundary issue in this Section. In \sec{polarization}, we prove that Convolutional Polar codes achieves the capacity of any binary-input memoryless symmetric channel (BMSC), and further that its error exponent is larger than that of Polar codes. Finite blocs-size numerical results comparing the error-correction properties of Convolutional Polar codes and Polar codes are presented in Sec.~\ref{sec:results}. We conclude and present some future directions in Sec.~\ref{sec:discussion}.

\section{Tensor networks and error correction}
\label{sec:TN}

Abstractly, we can view a gate, such as a CNOT, as a tensor $A^{\alpha \beta\gamma\ldots}$ with a certain number of indices denoted $\alpha, \beta, \gamma,\ldots$, each taking values in a finite set, that we will assume henceforth to be $\mathbb{Z}_2$.  The number of indices is the rank of the tensor. For instance, the CNOT gate is a rank-four tensor $N^{\alpha\beta\gamma \delta}$ with  indices $\alpha$ and $\beta$ representing the two input bits and $\gamma$ and $\delta$ representing the two output bits, and  the value of the tensor  given by $N^{\alpha\beta\gamma\delta} = 1$ if $\gamma = \alpha$ and $\delta = \alpha\oplus\beta$, and $N^{\alpha\beta\gamma\delta} = 0$ otherwise. We can graphically represent a tensor as a vertex and its indices as edges, with the degree of the vertex equal to the rank of the tensor. In that setting, an edge linking two vertices represents a tensor contraction defined by the following equation
\begin{equation}\includegraphics[width=6.5cm]{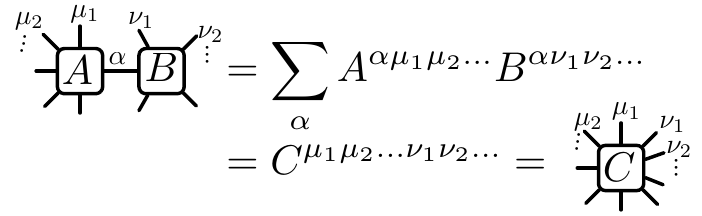}.\label{eq:TNC}\end{equation}
Tensor contraction is thus a very natural generalization of matrix multiplication. It follows from this definition that a (closed) graph represents a tensor network (TN) with all edges contracted, and hence a scalar. A graph with one open edge (an edge with one end not connected to a vertex) represent a tensor network with one uncontracted index, and thus a vector; and so forth. Such objects, and particularly their graphical representation, is also called a \emph{factor graph}, where some vertices may be identified as variables. \emph{Normal} factor graphs~\cite{FV11} provide a better analogy to open tensor networks.

\begin{figure}[t]
\centering
\includegraphics[width=\doublecolumnwidth]{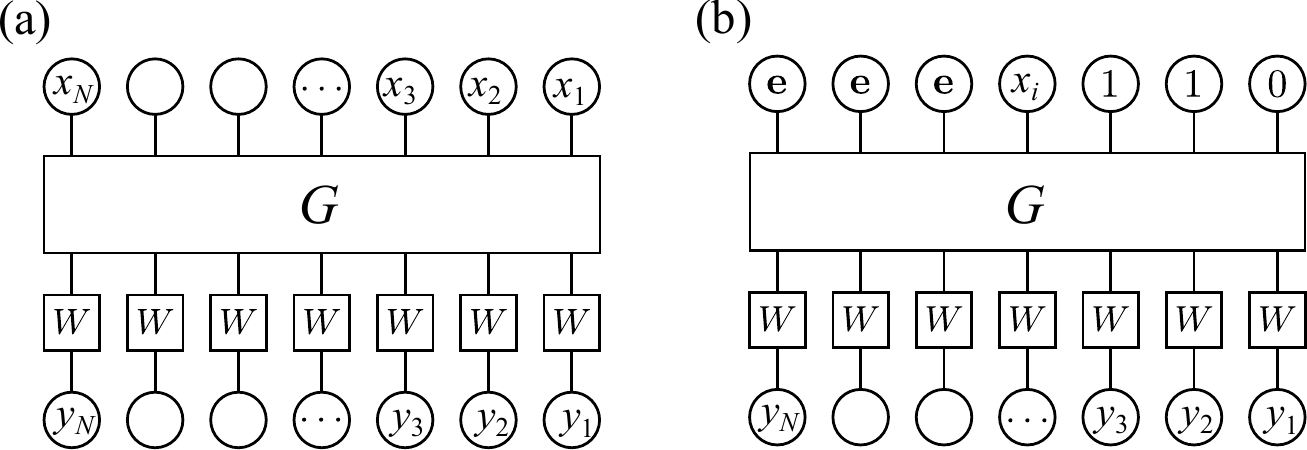}
\caption{(a) A simple TN diagram (or factor graph) of the generic decoding problem. The $N$ input bits $x_i$ are a combination of $k$ data bits and $N-k$ frozen bits, which are passed through the encoding circuit $G$. Given the measurements $y_i$ and the symmetric noise model $W$, we wish to determine the most likely configuration of data bits. The unnormalized probability $P(\mathbf{x}|\mathbf{y})$ is given by contracting the above TN, but it is not feasible to repeat for all $2^k$ possible inputs. (b) The successive cancellation decoder iteratively determines input bits in a right-to-left order. To determine the relative probabilities of bit $i$, we freeze the bits to the right using prior knowledge, while remaining completely ignorant about the states to the left, where $``{\bf e}"$ represents the uniform mixture $(1,1)$.  \label{fig:decoder}}
\end{figure}

Viewing the encoding circuit of a code --- such as a Polar code encoding circuit shown in \fig{circuits} (a) --- as a TN enables us to recast the decoding problem as a TN contraction problem. An  encoding circuit $G$ is a rank-$2N$ tensor, with $N$ indices representing $N$ input bits and $N$ indices representing $N$ output bits, where some of the input bits are fixed (frozen) to 0. A single-bit channel $W$ is a stochastic matrix, and hence a rank-two tensors. Finally, we can represent the probability distribution over the value of a single bit as a rank-one tensor, with the tensor $``0" = (1,0)$ representing the bit value 0 and tensor  $``1" = (0,1)$ representing the bit value 1. Given these, the probability of the input bit string  $\mathrm{x} = (x_1,\dots,x_N)$ given the observed output $\mathrm{y} = (y_1,\dots,y_N)$ can be represented as the TN shown in \fig{decoder}~(a). 

In general, not all TNs can be efficiently contracted. Referring to \eq{TNC} where tensor $A$ has rank 6 and tensor $B$ has rank $5$, we see that the tensor $C$ resulting from their contraction has rank $6+5-2 = 9$. Thus, while tensor $A$ is specified by $2^6$ entries and tensor $B$ is specified by $2^5$ entries, tensor $C$ contains $2^9 \gg 2^6+2^5$ entries. A TN composed of bounded-rank tensors (e.g., a circuit with only two-bit gates) can be specified efficiently. However, the tensors obtained at intermediate steps of the TN contraction schedule can be of very high rank $r$, and so its contraction will produce an intractable amount of data $2^r$. The contraction schedule that minimizes the intermediate tensor rank defines the treewidth of the graph, so generally the cost of contracting a TN is exponential with its treewidth \cite{MS08a}. 

This implies that encoding circuits that produce TNs with finite treewidth can be efficiently decoded. This is the case for instance of Convolutional codes \cite{JZ99a}, whose corresponding TN is simply a chain, and therefore have a constant treewidth. However, it can sometimes be possible to efficiently decode even when the encoding circuit has a large treewidth by making use of special circuit identities that simplify the TN. An example is provided by the fact that a CNOT gate with a 0 entry on the controlled bit is equivalent to the identity --- see \fig{identities}~(b) for the corresponding graphical identity. The combination of such circuit identities provides a powerful graphical calculus that can be used to contract highly complex TNs. In particular, Arikan's sequential cancellation decoding can be recast in this graphical calculus as an efficient TN contraction that relies precisely on the identities in \fig{identities} (this is described in more detail in \sec{C-decoding}). 

Graphical calculus is commonly employed in quantum physics, starting with Feynman diagrams for quantum electrodynamics, to quantum circuit representation of quantum computations. More recently, a graphical calculus was developed for the representation of quantum many-body states \cite{V03a,VC04a,SDV06a,Vid05a,EV12b}. The quantum state of a system comprising $N$ qubits or spin-1/2 particles, for example, is a $2^N$-dimensional vector, so its specification requires an exponential amount of data. A vector with $2^N$ components can be viewed as  a rank-$N$ tensor with binary indices. Thus, by restricting  to  tensors that are obtained from the contraction of polynomially many bounded-degree tensors, we reduce the amount of data required to specify a quantum state from exponential to polynomial. Then, the evaluation of physical quantities of interest (energy, magnetization, etc.) amounts to the problem of contracting the corresponding tensor network. 

In \cite{FP142}, we demonstrated the equivalence between several TN families developed in the context of quantum many-body physics and encoding circuits of various classical and quantum error correcting codes. In particular, the computational techniques developed in physics and coding theory to evaluate quantities of interest (e.g. magnetization in physics, bit likelihood in coding theory) are often identical. Of particular interest here are TNs called branching MERA, which were introduced in physics both as a conceptual and numerical tool to understand the properties of potentially highly entangled physical systems which exhibit separation of degrees of freedom at low energies \cite{EV12b}. The graph representing this TN has a richer structure than the encoding circuit of Polar code, but yet it remains efficiently contractible. This is the key observation which enables us to define a generalization of Polar codes.

As a side remark, the graphical calculus of TNs has been discovered in numerous scientific fields were it is known under  different names. Beyond the examples from physics already mentioned, these include factor graphs, Markov random fields, Bayesian networks, partition functions, and trace diagrams in coding and information theory~\cite{AM00,YFW03,M03,MM09}; stochastic Petri nets from biology; chemical reaction networks; artificial neural networks, connectionist systems, and Boltzmann machines from artificial intelligence, and so on~\cite{M03,PL79, PL85, BE14}. In this last setting, the use of convolutional structures has led to a burst of applications including deep convolutional neural networks.  Unifying all these are the mathematical field of network theory and category theory~\cite{Pen71, BCJ11}. Each of the above are examples of symmetric monoidal categories, and there exist rigorous proofs of the equivalence of the graphical notation and manipulations with traditional algebraic relations~\cite{L95}. Thus the circuit and tensor network diagrams here should be considered as formal equations and expressions. The study of category theory concerns itself with identifying mathematical results known in one setting and formally applying them to another --- much in the same spirit of this work.

\section{Encoding and decoding}
\label{sec:geo}

In this Section we describe the encoding circuit of Convolutional Polar codes, focusing on its relationship to Arikan's Polar code, followed by an analysis of its linear encoding matrix, before finishing with an efficient algorithm for successive cancellation decoding.

\subsection{Encoding circuit}

\begin{figure}[t]
\centering
\includegraphics[width=1.6\doublecolumnwidth]{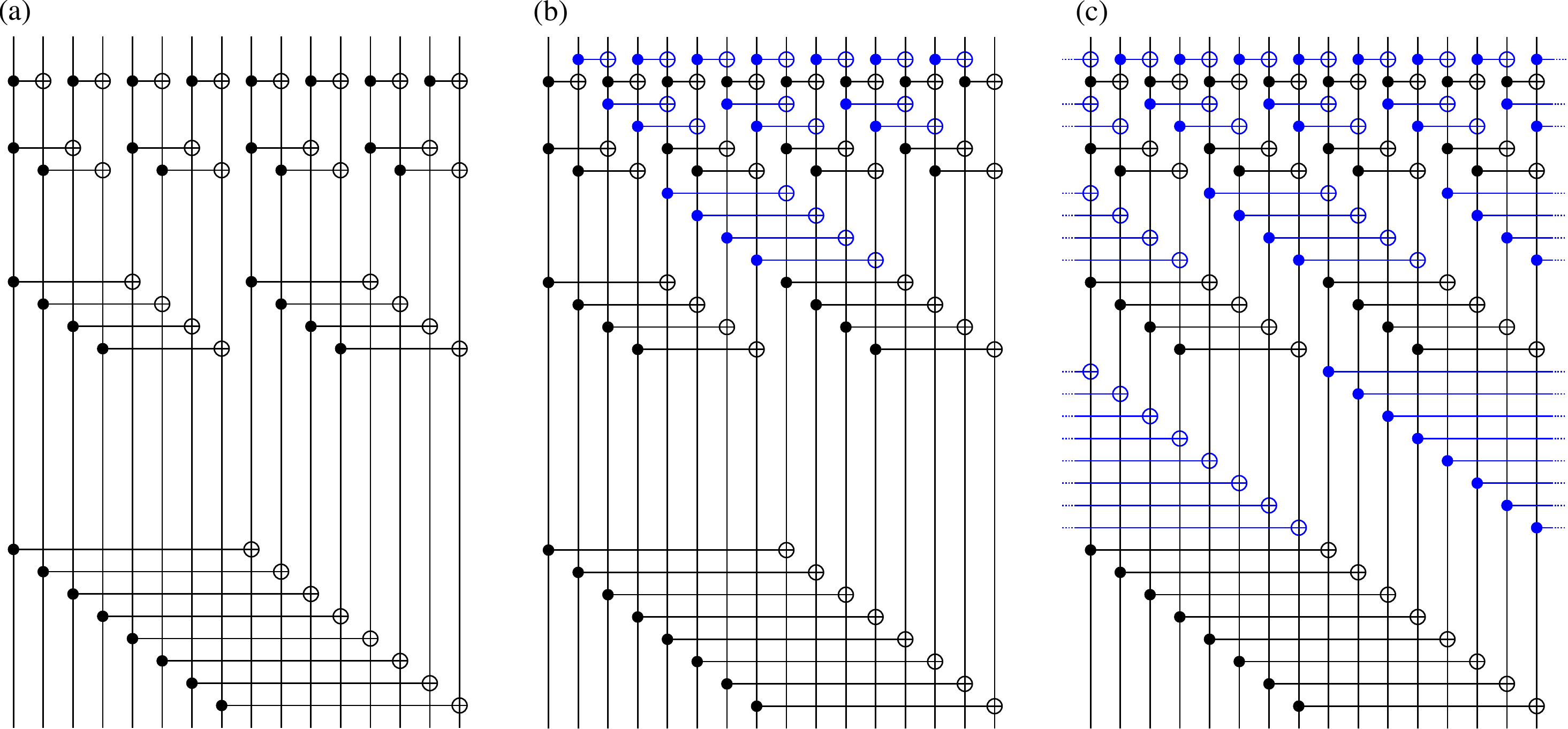}
\caption{The encoding circuits of (a) the Polar code,  (b) the Convolutional Polar code with open boundaries, and (c) the Convolutional Polar code with periodic boundaries, for $2^4 = 16$ sites. The logical bits enter into the top of the circuit, and the encoded message appears at the bottom. The Polar code contains half of the gates of the periodic Convolutional Polar code. The extra gates are highlighted in blue in (b) and (c), while the dots represent periodic boundaries.  \label{fig:circuits} }
\end{figure}

The definition of a branching MERA code relies purely on the \emph{topology} of the encoding circuit. For the special case of Polar and Convolutional Polar codes, these definitions are:

\begin{definition} \label{def}
The encoding circuit for the Polar or the Convolutional Polar code over $N = 2^n$ bits (labelled $1\dots N$) is constructed recursively using the following rules. If $N = 1$, we begin with the trivial code. An encoding circuit for a code of size $N = 2^n$ can be constructed from those of two codes of size $2^{n-1}$ by interleaving the logical channels and adding the following two-bit gates to the \emph{beginning} of the circuit:
\begin{enumerate}
\item For the Convolutional Polar code, we begin with a layer of $2^{n-1}$  CNOT gates, where the $i$th gate connects bits $2i$ to $2i+1$ (for $i \ne 2^{n-1}$). The final gate connects bit $N$ with bit $1$. If this gate is non-trivial, the Convolutional Polar code is said to periodic (otherwise it is described as non-periodic or having open boundary conditions).
\item For either the Polar or Convolutional Polar code, we next apply $2^{n-1}$  CNOT gates, where the $i$th gate connects bits $2i-1$ to $2i$.
\end{enumerate}
\end{definition}
The $N=16$ Polar and Convolutional Polar encoding circuits are depicted on \fig{circuits}. 

Branching MERA codes include generalizations of this definition, for instance to higher dimensional fields, by increasing the number of gates at each layer of recursion, or by fusing $k > 2$ codes at each layer using $k$-bit gates. In many of these cases, this opens up the possibility of using nonlinear gates for the kernel. We claim our results regarding efficient encoding and decoding hold for all of these extensions, though we only prove them for the geometry in Def.~\ref{def} corresponding to the Polar and Convolutional Polar codes. It should be noted that not all branching MERA codes, according to the above definition, exhibit channel polarization --- that will depend on the choice of gates used.

For the remainder of this paper we will focus on just these two cases, in their respective cases referring to them simply as ``the Polar code' and `the Convolutional Polar code''.  From \sec{polarization} onward, we will further specialize to the open-boundary Convolutional Polar code. It follows straightforwardly that encoding is efficient:
\begin{proposition}
The encoding circuit for a (Convolutional) Polar code over $N$ bits can be applied with a computation cost scaling as $N \log_2 N$, and if we allow parallel processing, only takes time scaling as $\log_2 N$.
\begin{IEEEproof}
The total number of two-bit gates to encode using the Polar code is simply $N/2 \log_2 N$. The (periodic) Convolutional Polar code has precisely twice as many gates introduced at each layer, making for a total of $N \log_2 N$. From the construction, it is clear that only one (Polar) or two (Convolutional Polar) layers of gates are applied every time the code size doubles, so the circuit depth is logarithmic: $\log_2 N$ for the Polar code and $2\log_2 N$ for the Convolutional Polar code.
\end{IEEEproof}
\end{proposition} 

\subsection{Linear encoding matrix}

For the binary codes studied here, we are naturally restricted to considering linear gates. We will now compare and contrast the linear encoding matrices for the Polar and Convolutional Polar codes.

\begin{figure}[t]
\centering
\includegraphics[width=0.6\doublecolumnwidth]{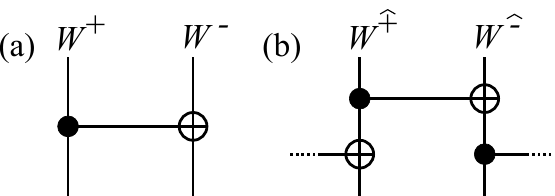}
\caption{The basic building blocks of (a) the Polar code, and (b) the Convolutional Polar code. The Polar code can be thought of as an attempt to strengthen the left logical channel at the expense of the right, under successive cancellation decoding. In the Convolutional Polar code, information from all input channels is spread out, from left to right and at different length scales (with periodic boundaries). A specific choice of successive decoding order  makes the leftmost channels less susceptible to noise than the right, on average. \label{fig:primitives} }
\end{figure}

The Polar code is based on the idea of polarization of the input channels into those that are almost noiseless or very noisy, under the successive cancellation decoder. The basic primitive of the Polar code is the CNOT gate, as depicted in Fig.~\ref{fig:primitives}~(a). The CNOT gate is a linear gate represented by the matrix
\begin{equation}
  G_2 = \left[ \begin{array}{cc} 1 & 1 \\ 0 & 1 \end{array} \right],
\end{equation}
and it acts to copy the data from the left input channel and add it to the right, essentially giving that data more opportunities to avoid corruption by the noise. Conversely, the data on the right logical channel may become obscured by uncertainty in the data from the left --- particularly as the successive cancellation decoder will not make that determination until later. In the Polar code, the gate is applied in parallel in a number of layers, and the encoding matrix for a single layer of parallel application becomes, for example, as $G_2 \otimes I \otimes I \otimes \dots$, where $I$ is the identity matrix. 

Here, we go beyond this picture by continuing to copy the data in a left-to-right fashion by using a second layer of CNOT gates. The primitive in Fig.~\ref{fig:primitives}~(b) is applied everywhere on the lattice, connecting all sites (in a periodic structure) so that every logical channel becomes the control bit of at least one CNOT. The goal is to more evenly and rapidly spread out the information, in the hope that the data is better protected from noise, as well as to increase the impulsive response of the encoder (and thus possibly the distance of the code).

In both cases, these primitives are composed (or concatenated) on different length scales --- with the distance spanned by the CNOT doubling at each `layer' of the code. In the original Polar code, it was observed that the channel polarization increased with each additional layer, with the large codes approaching the capacity of binary symmetric channels. The linear encoding matrix for the $i$th layer of a $N=2^n$ code becomes $A_i = I_2 ^ {\otimes i-1} \otimes G_2 \otimes_2 I^{n-i}$. Thus, we can most easily express the Polar code encoding  matrix in a recursive form 
\begin{align}
G_{2^n} &= (G_{2^{n-1}}\otimes I_2) \cdot (I_{2}^{n-1}\otimes G_2) \label{eq:recursiveP} \\
&= (G_{2^{n-1}}\otimes I_2)\cdot
\left(
\begin{array}{cccccccc}
1&1&0&0&0&0&0&0\\
0& 1& 0&0&0&0&0&0\\
0&0&1&1&0&0&0&0\\
0&0&0&1&0&0&0&0\\
0&0&0&0&1&1&0&0\\
0&0&0&0&0&1&0&0\\
0&0&0&0&0&0&1&1\\
0&0&0&0&0&0&0&1\\
\end{array}\right).
\end{align}
(where the last equality is illustrated for $n=3$) which leads to the simple solution 
\begin{equation}
   G_{2^n} = G_2 \otimes G_2 \otimes G_2 \otimes \dots = \left[ \begin{array}{cc} 1 & 1 \\ 0 & 1 \end{array} \right] ^ {\otimes n}.
\end{equation} 
Although the layers commute, we note that we have fixed the order of the layers as a matter of convention in Def.~\ref{def}. This ordering can be seen in Fig. \ref{fig:circuits}~(a). 

With the Convolutional Polar code, we continue to compose layers together, spreading out the information on exponentially growing length scales. 
Each layer contains the Polar code gates with encoding circuit $A_i$ and an addition layer given by $A_i$ translated by $2^{i-1}$ sites. We thus see that \eq{recursiveP} is replaced by
\begin{align}
G_{2^n} &= (G_{2^{n-1}}\otimes I_2) \cdot \left[(I_2^{n-1}\otimes G_2)S_{2^n}(I_2^{n-1}\otimes G_2)S_{2^n}^T\right] \label{eq:recursive}\\
&= (G_{2^{n-1}}\otimes I_2)\cdot
\left(
\begin{array}{cccccccc}
1&1&1&0&0&0&0&0\\
0& 1& 1&0&0&0&0&0\\
0&0&1&1&1&0&0&0\\
0&0&0&1&1&0&0&0\\
0&0&0&0&1&1&1&0\\
0&0&0&0&0&1&1&0\\
1&0&0&0&0&0&1&1\\
1&0&0&0&0&0&0&1\\
\end{array}\right),
\end{align} 
where $S_k$ is the cyclic permutation matrix of $k$ elements, e.g. for $k=8$, 
\begin{equation}
S_8 = \left(
\begin{array}{cccccccc}
0 & 0 & 0 & 0 & 0 & 0 & 0 & 1\\
1&0&0&0&0&0&0&0\\
0&1&0&0&0&0&0&0\\
0&0&1&0&0&0&0&0\\
0&0&0&1&0&0&0&0\\
0&0&0&0&1&0&0&0\\
0&0&0&0&0&1&0&0\\
0&0&0&0&0&0&1&0\\
\end{array}\right)\label{eq:translation}.
\end{equation}Comparing Eqs.~(\ref{eq:recursiveP}) and (\ref{eq:recursive}), the key distinction between the two codes is that the matrix by which $(G_{2^{n-1}}\otimes I_2)$ gets multiplied is not block diagonal. Thus, unlike the Polar code, the composed layers do not commute, and efficient decoding will rely crucially on the ordering.  While the Polar code encoding circuit is idempotent ($G_N^2 = I$), the Convolutional Polar code is not. We have found numerically that $G_N^N = I$; that is, applying the $N$-bit encoding circuit $N$ times is equivalent to the identity operation.

\begin{figure}[t]
\centering
\includegraphics[width=0.33\doublecolumnwidth]{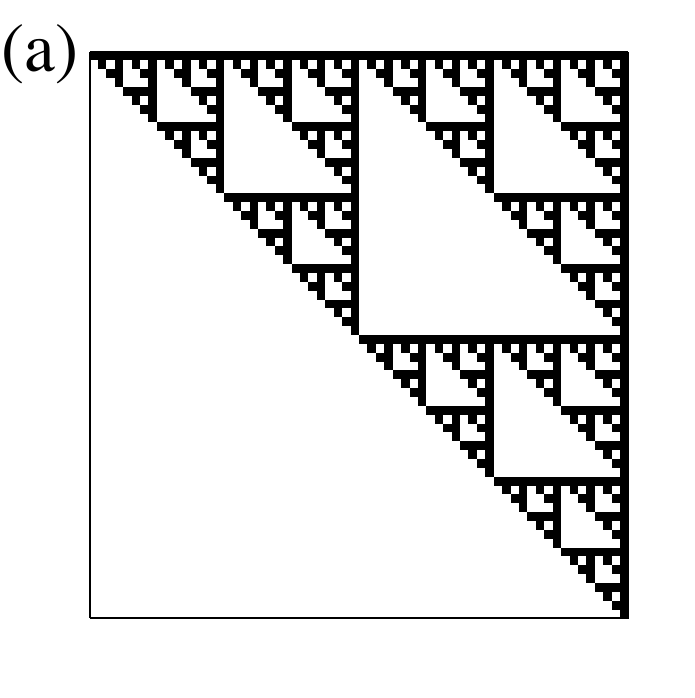}\includegraphics[width=0.33\doublecolumnwidth]{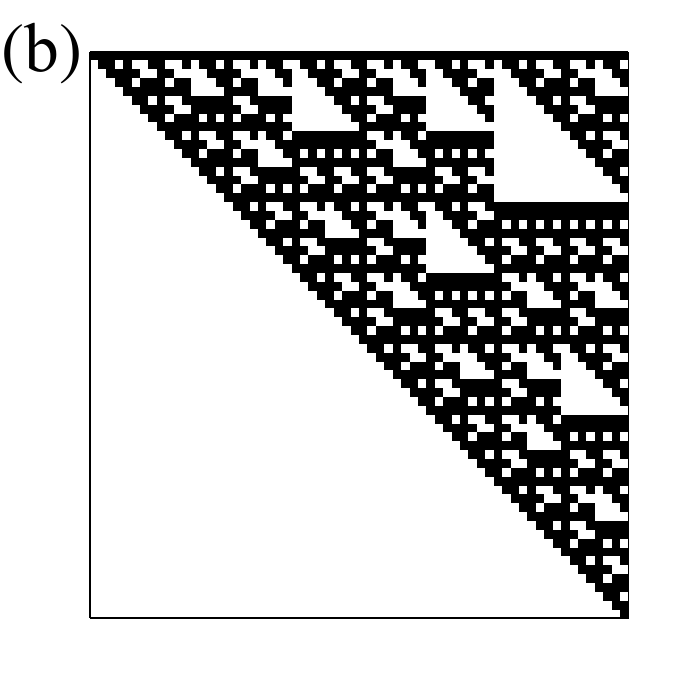}\includegraphics[width=0.33\doublecolumnwidth]{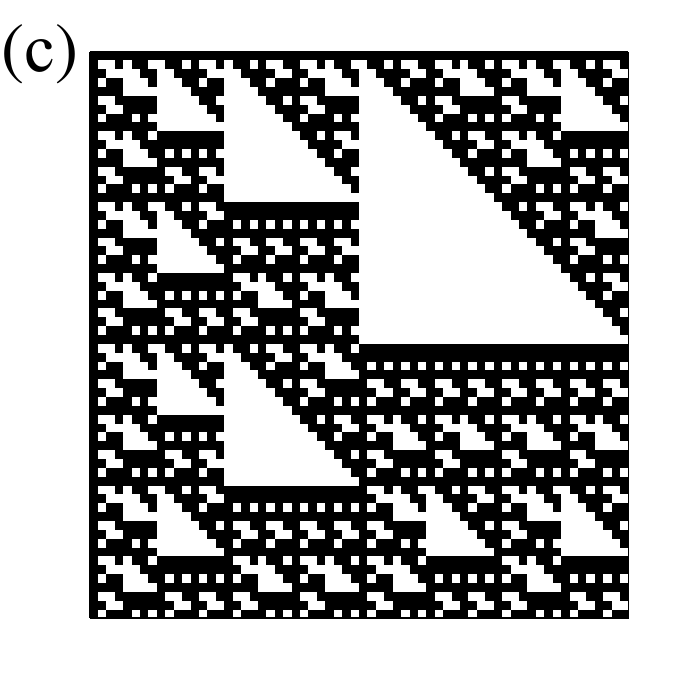}
\caption{Graphical representation of the linear encoding matrix for 64-bit codes for (a)~the Polar code, (b)~open-boundary Convolutional Polar code, and (c)~periodic-boundary Convolutional Polar code. Here white represents matrix entries of 0 and black represents 1. \label{fig:matrices} }
\end{figure}

\subsection{Boundaries}

We have defined Convolutional Polar codes with periodic or open boundary conditions. The open boundary code is obtained from the periodic boundary code simply by removing the gate which connects bit $2^n$ to bit $1$ in the $n$-th level of polarization.  Open and periodic boundary Convolutional codes are contrasted for $N=16$ at Fig.~\ref{fig:circuits}~(b)~\&~(c).  The difference between the two can be  summarized by presence or absence of a 1 in the upper right corner of the translation matrix $S$, c.f., \eq{translation}. The encoding matrices are compared in Fig.~\ref{fig:matrices}, along with the encoding matrix of the Polar code. 

While such boundary conditions lead to two distinct encoding circuits, we note that under successive cancellation decoding, the periodic and the open boundary Convolutional Polar codes become almost identical. 
This is because the additional gates in the periodic code mainly copy information from early bits onto later bits and therefore, when invoking the successive cancellation decoder for one of the early bits, this additional information is not available yet and cannot help as side information for decoding that bit. Similarly, when we attempt to decode the later bit, we will already know the earlier ones perfectly and therefore we can simply add its value again and therefore deterministically remove the effect of the gate. Because the polarization layers of the Convolutional Polar code do not commute with one another, some of the periodic gates appearing in later polarization stages cannot be removed. This is illustrated on \fig{decoders} (b) and (c). In the rest of this manuscript, we choose to focus on the open-boundary code, but almost everything goes through using the periodic one. 

The boundary causes an important departure from Polar codes in the polarization analysis of Convolutional Polar codes. The basic principle of Polar codes lies in the Polar transform which synthesizes two channels $W^-$ and $W^+$ from two independent copies of the channel $W$ by appropriate combination via a CNOT gate. This can in principle also be done with two different channels, in this case we denote the equivalent of $W^\pm$ by $\langle W_1, W_2 \rangle^{\pm}$. Similarly, each polarization layer of the Convolutional Polar code synthesizes two types of channels, c.f., \fig{primitives}, that we denote  
\begin{align}
W^{\widehat{-}} = \langle \langle W,W\rangle^{-} , \langle W,W\rangle^{+} \rangle^{-} \\
W^{\widehat{+}} = \langle \langle W,W\rangle^{-} , \langle W,W\rangle^{+} \rangle^{+}.
\end{align}
These channel types define the majority of channel combinations found in Convolutional Polar codes, and we will refer to them as {\em bulk} transformations. In addition to these bulk transformations, other types of transformations occur near the boundary of the code. These differences are caused by gates which were removed when imposing open boundary conditions or when choosing a starting point for the successive cancellation decoder in the periodic case. These include combinations such as $\langle W,\langle W,W\rangle^+\rangle^\pm$, and so forth.  More precisely, in the open-boundary code, these {boundary} channels in the $l$-th polarization step are those with index $i \leq 2^{l}$ or $i\geq 2^{n-l}$. Note that in each step $l$ the boundary channels are a subset of the ones in the $l+1$th step, therefore their numbers don't add up. More precisely, the fraction of boundary channels in step $l$ is 
\begin{equation}\label{eq:fraction}
2^{l+1-n}.
\end{equation}
This will be helpful in the polarization proof of \sec{polarization}.

\subsection{Decoding algorithm}
\label{sec:C-decoding}

It is not immediately obvious that the Polar or Convolutional Polar codes can be efficiently decoded. One of Arikan's achievements was to realize that the decoding problem simplifies significantly under a successive cancellation scheme. With this decoder, the goal is to determine a single bit at the time, moving from right-to-left, by assuming complete ignorance of the input bits to the left, and total confidence in the value of the input bits to the right (either because they are frozen in the code, or because we have decoded those bits already). In Fig.~\ref{fig:decoder}, we wrote down the central probability density calculations required for a generic decoder and the successive cancellation decoder in a simple tensor network diagram.

A generic, optimal decoder will locate the codeword with maximal likelihood --- that is, the most probable input $\mathrm{x} = (x_1,\dots,x_N)$ given the observed output $\mathrm{y} = (y_1,\dots,y_N)$, error model $W$, and the set of frozen bits $A^c$:
\begin{equation}
   \max_{\mathrm{x}} P(\mathrm{x} | \mathrm{y}, \{x_k,  \forall k \in A^c\}).
\end{equation}
However, for many codes determining the most probable codeword exactly is a hard problem, and a range of (usually iterative) approximations are used. The successive cancellation decoder begins with the rightmost non-frozen bit at position $i$, and determines its value by maximizing
\begin{equation}
   \max_{x_i} P(x_i | \mathrm{y}, \{x_k,  k = i+1,\dots,N \}).
\end{equation}
For the purpose of the above calculation, the bits to the left of $i$ (i.e. $1,\dots,i-1$) are considered unknown, even if they are frozen bits. In this sense, successive cancellation is not an optimal decoder, because it does not take advantage of all the available information at every step. It then proceeds to the next non-frozen bit, and so on, until the complete message has been determined.

The tensor networks depicted in Fig.~\ref{fig:decoder} are calculations on bit probability distributions. There, we have used the tensor notation  $``0" = (1,0)$ and   $``1" = (0,1)$ as above, and have introduced the (un-normalized) uniform distribution ${\bf e} = (1,1)$ to represent a bit of which we have no knowledge. Furthermore, receiving the bit $y_i$ transmitted through a channel $W$ implies a probability distribution on the output bit $p_i$ that is given by Bayes' rule. For symmetric channels $W$ 
this is simply $p_i \propto W^T y_i$.

\begin{figure}[t]
\centering
\includegraphics[width=1\doublecolumnwidth]{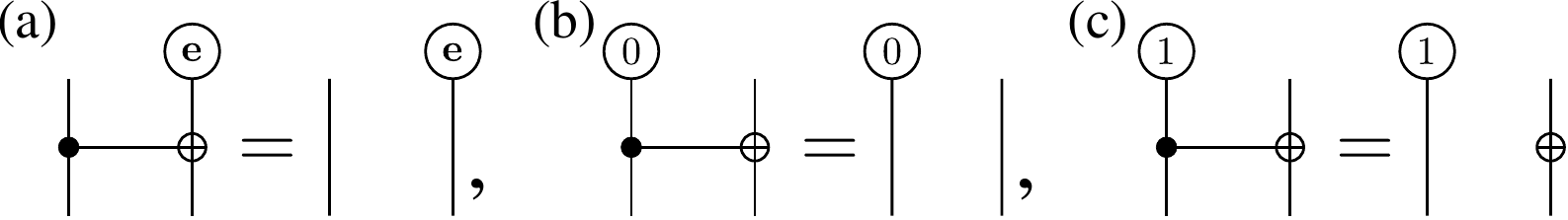}
\caption{Three basic circuit identities relating how the CNOT acts on probability distributions. These represent \emph{every} product input distribution that results in an uncorrelated, product output distribution. We must use these to simplify the tensor network contraction required for the successive cancellation decoder, for both the Polar and Convolutional Polar codes.  \label{fig:identities} }
\end{figure}

The CNOT gate has a very simple action on some states, such that it does not introduce any correlations to the joint distribution. In Fig.~\ref{fig:identities} we detail {\em circuit identities} that define the action of CNOT on the distributions $``0"$, $``1"$, and $\bf e$. Generically, similar identities hold for all reversible (i.e. deterministic, one-to-one) gates --- even non-linear gates. The proofs that follow only rely on the ability to calculate the action of the gate, and that fully random inputs lead to fully random outputs.

\begin{figure}[t]
\centering
\includegraphics[width=1\doublecolumnwidth]{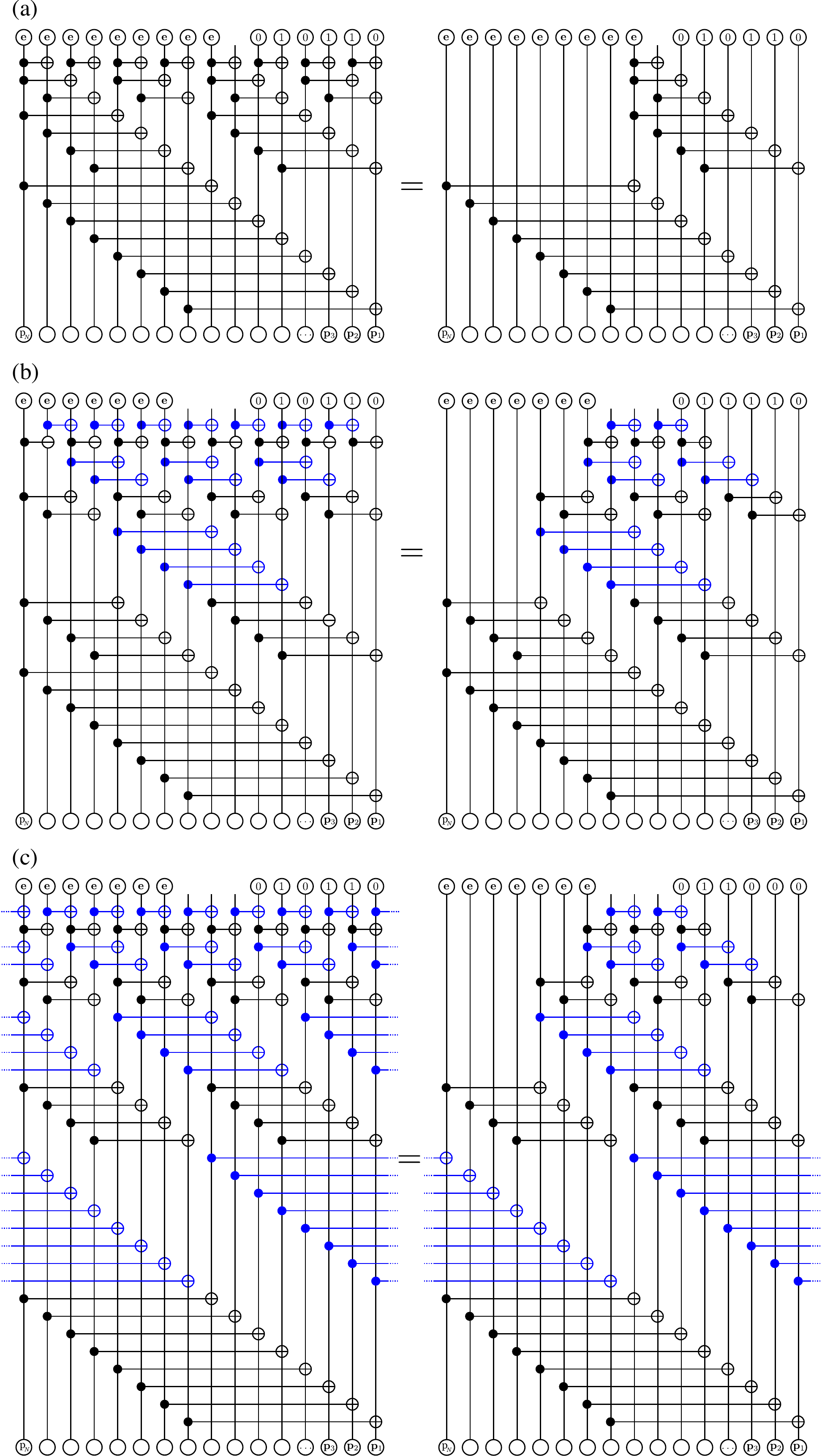}
\caption{The simplified tensor networks for successive cancellation decoding of (a) the Polar code, (b) the open-boundary Convolutional Polar code, the periodic-boundary Convolutional Polar code using the identities illustrated in Fig.~\ref{fig:identities}. These tensor networks contain open (non-contracted) indices as proxies for any tensor that could be placed at these locations. For the Convolutional Polar code, it is natural to determine the joint probability distribution of three neighbouring bits, because of the iterative scheme in Fig.~\ref{fig:contractions}~(b). All three tensor networks can be contracted from the bottom-up in a time \emph{linear} in the total number of bits, $N$. \label{fig:decoders} }
\end{figure}

Applying these identities to the Polar and Convolutional Polar codes results in a vast simplification. In fact, \emph{most} of the CNOT gates are removed, as illustrated at \fig{decoders} (a) for the Polar code and (b) for the Convolutional Polar code. The number of remaining gates drops from $\mathcal{O}(N \log_2 N)$ to $\mathcal{O}(N)$. We re-derive this for the Polar code for clarity:
\begin{lemma}
To decode the $i$th bit using successive cancellation on the Polar code of $N$ bits, all but $N-1$ gates can be removed from the tensor network diagram.
\begin{IEEEproof}
During successive cancellation, bits $1,\dots,i-1$ are unknown with probability distribution $\mathbf{e}$, while bits $i+1,\dots,N$ have known values described by the probability distributions ``0'' or ``1''. We focus first on the top `layer' of gates. If $i$ is odd (or even), then the gates connecting bits $1,\dots,i-1$ (or $1,\dots,i-2$) all have input $\mathbf{e}$, and therefore can be removed without effect using the identity in Fig.~\ref{fig:identities}~(a). The bits $i+2,\dots,N$ (or $i+1,\dots,N$) are fully determined and their gates can be removed by manipulating the bit values according to the action of the gate, as in Fig.~\ref{fig:identities}~(b,c). In this layer only one gate remains, connecting bits $i,i+1$ (or $i-1,i$). 

The encoding circuit was constructed recursively according to Def.~\ref{def}. The remaining gates are the single one indicated in the top layer, plus those of the Polar code of $N/2$ bits over the odd sites, and an identical code over the $N/2$ even sites. Within each `sub-code', the upper CNOT has exactly one output at position $i$ (or either $i-1$ or $i+1$, depending on whether $i$ it is even or odd), while the bits to the left are unknown and the bits to the right are known. Therefore, we arrive back to the original situation. Within each sub-code, all but one CNOT can be removed. For every layer we descend, the number of sub-codes doubles. For $N = 2^n$ there are $n$ layers, and we have $\sum_{i=0}^{n-1} 2^i = 2^n - 1 = N-1$ gates remaining.
\end{IEEEproof}
\end{lemma}

Although the encoding circuit for the Convolutional Polar code is more complicated, it is constructed iteratively in the same manner. For the Polar code we saw that probing a single bit in the code resulted in effectively probing a single bit in two smaller codes, one layer below. This in not the case for the Convolutional Polar code --- it turns out that probing a three-bit distribution in one layer can be mapped to probing three-bit distributions in two smaller codes below.
\begin{lemma}
To decode bits $i,i+1,i+2$ using successive cancellation on the Convolutional Polar code of $N$ bits, strictly less than $5N$ gates remain in the tensor network diagram.
\begin{IEEEproof}
During successive cancellation, bits $1,\dots,i-1$ are unknown with probability distribution $\mathbf{e}$, while bits $i+3,\dots,N$ have known values described by the probability distributions ``0'' or ``1''. The top layer consists of two rows of CNOT gates. In the upper row, only two CNOT gates are connected to bits $i,i+1,i+2$. The remaining gates can be removed using the identities in Fig.~\ref{fig:identities}. These two CNOT gates necessarily connect to three CNOT gates in the second row, which are offset by one site. All the remaining gates in the second row can also be removed using the same identities. There are a total of 5 gates in the top layer.

The remaining gates in this layer connect to a total of 6 bits, three of which are on odd sites and three on even sites. Due to the construction of the Convolutional Polar code in Def.~\ref{def}, we can split the odd and even sites into two effective Convolutional Polar codes, each of which have three bits connected to the CNOTs above. The bits to the left are unknown, and the bits to the right are fully determined, and so the process iterates. Another 5 gates will remain in both the `odd' and `even' sub-code.

For the bottom two layers, the number of bits in each code becomes fewer than 6, so the iteration ends. At the second-bottom layer, there are only four gates total, while there are just two in the lowest --- in both cases less than five. The total number of gates is upper-bounded by $5(N-1)$.
\end{IEEEproof}
\end{lemma}

Now that the tensor network diagram has been simplified, it  remains to be shown that it can be contracted efficiently. While the gate cancellation is done from top-to-bottom, the tensor network contraction is done from bottom-to-top. 

\begin{figure}[t]
\centering
\includegraphics[width=\doublecolumnwidth]{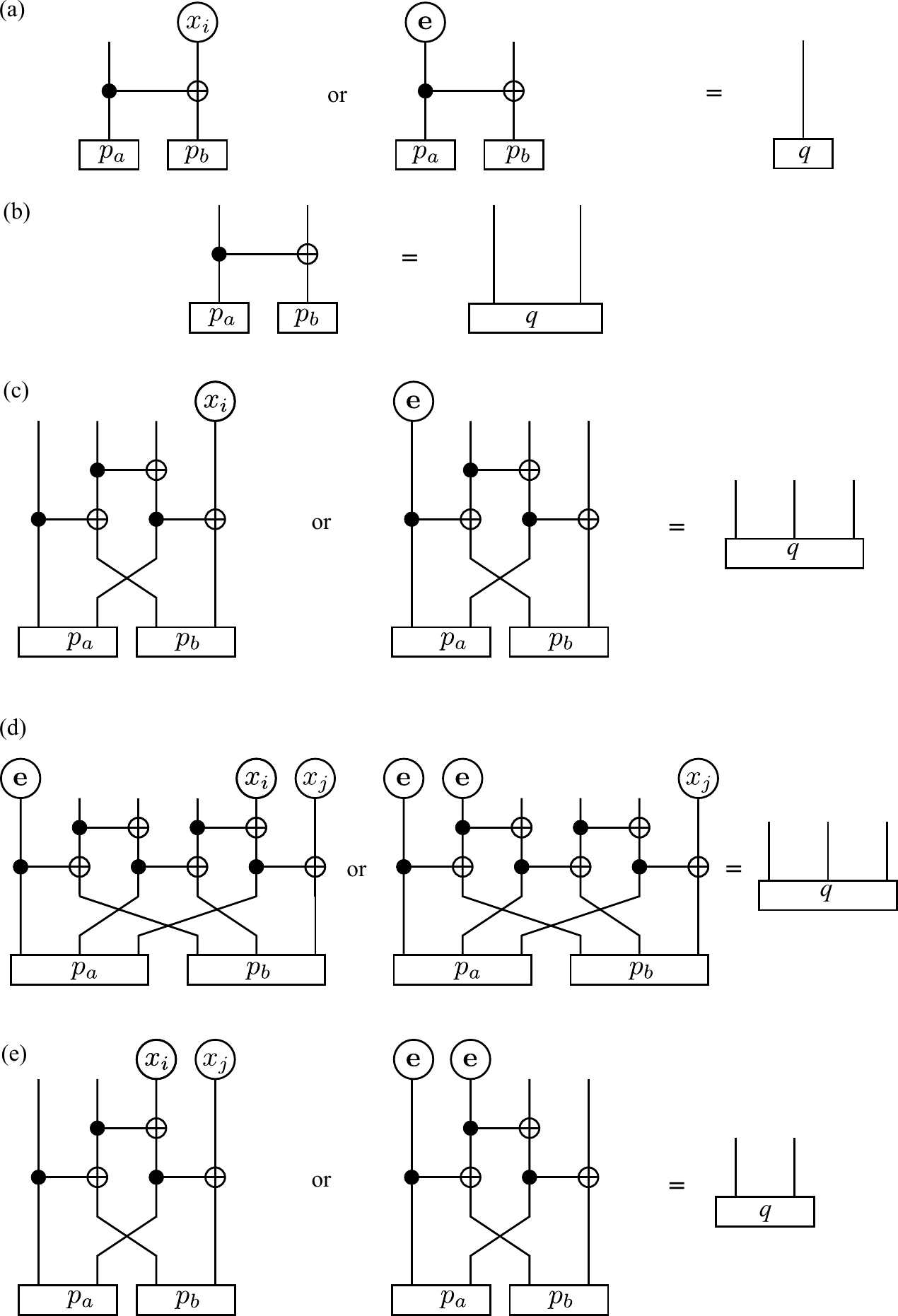}
\caption{The basic contractions required to decode (a) the Polar code and (b--e) the Convolutional Polar code. As opposed to the identities of \fig{identities}, these contraction identities apply to arbitrary tensors. The tensor $q$ depends on the specific tensor network being considered.  (a)~One of these basic transformations is selected at each `layer' of the Polar code, depending on the input bit targeted. (b)~The Convolutional Polar code always combines two bits at the lowest layer. (c)~After this, the distributions are combined to a 3-bit distribution. (d) The 3-bit to 3-bit transformation is a natural fixed point of the branching MERA~\cite{EV12b}. (e)~When the input bit is near the boundary, the lower layers may make several 2-bit to 2-bit transformations before moving to the 3-bit ones in (c,d).\label{fig:contractions} }
\end{figure}
To see how this is done, it is useful to highlight a few {\em contraction identities} shown at \fig{contractions}. In contrast to the circuit identities of \fig{identities} the contraction identities of \fig{contractions} are purely graphical in that they do not depend on the nature of the gates being contracted but only on the underlying graph structure. For instance, the CNOTs used in these contraction identities could be replaced by any other rank-four tensor and preserve the identities. Applying these graphical identities repeatedly starting from the bottom of the diagrams of \fig{decoders} and working our way up yields an efficient contraction schedule. We will first re-prove this for Polar code for clarity, and extend this to the Convolutional Polar code.

\begin{lemma}
A single step of successive cancellation decoding in the $n$-bit Polar code has computational cost linear in $N$.
\begin{IEEEproof}
Assume we are decoding bit $i$. At the bottom of the tensor network diagram, we begin with $N$ single-bit probability distributions (given by the error model and received bits). These are connected via $N/2$ CNOT gates in the lowest layer of the encoding circuit, where, if $i \le N/2$ the right bit is already determined, or if $i > N/2$ the left bit is completely unknown (in state $\mathbf{e}$). We can apply the appropriate intermediate step in Fig.~\ref{fig:contractions}~(a) to obtain $N/2$ single-bit distributions, moving up one layer in diagram. Each application of Fig.~\ref{fig:contractions}~(a) has a fixed cost, and here we performed it $N/2$ times.

Following this, we have returned to a self-similar situation but with only $N/2$ total bits remaining, and so we iterate the procedure. At each layer, the number of bits halves, and we apply the appropriate contraction from Fig.~\ref{fig:contractions}~(a) (at the $l$th layer we determine this according to the $l$th bit of the binary decomposition of $i-1$). At the top, we are left with the probability distribution of the logical bit, from which we can make a determination of its value. The total number of contraction steps is $N/2 + N/4 + \dots + 1 = N-1$.
\end{IEEEproof}
\end{lemma}

The Convolutional Polar code is decoded in a similar fashion, but requires multiple-bit probability distributions.
\begin{lemma}
A single step of successive cancellation decoding in the $N$-bit Convolutional Polar code has computational cost linear in $N$.
\begin{IEEEproof}
Assume we are decoding bits $i,i+1,i+2$. We break the problem into three steps: the lowest layer, the second-lowest layer, and the remainder.

At the bottom, we begin with single-bit distributions. However, more CNOT gates lie above these and so we must combine the distributions under the action of the CNOT according to Fig.~\ref{fig:contractions}~(b). Each has fixed cost and this occurs $N/2$ times.

At the second-lowest layer, in the bulk (see below for the boundary), these two-bit distributions are combined into three-bit distributions, according to Fig.~\ref{fig:contractions}~(c). Which diagram is relevant depends on the value of $i$ (placing $i$ into binary form will make this readily apparent). This will occur $n/4$ times, and although the cost is greater than the previous layer, the cost of each contraction is fixed.

At the remaining layers, in the bulk, we combine a pair of three-bit distributions to obtain a three-bit distribution one level higher using the steps in Fig.~\ref{fig:contractions}~(d). At each level, the number of distributions halves. This occurs a total of $N/8 + N/16 + \dots + 1$ times, resulting in $N/4 - 1$ contraction steps. Including the bottom two layers, the total number of steps is $N-1$

Contractions involving boundary channels have fewer CNOT gates. But since the above proof relies only on graphical circuit identities, it remains valide when a CNOT is substituted by an identity gate. More concretely, the third step might not be reached immediately; instead, after the second-lowest layer we remain with a two-bit distribution using the contraction in Fig.~\ref{fig:contractions}~(e). At some point, the distribution will widen to three bits according to the pattern of remaining CNOT gates, as in Fig.~\ref{fig:contractions}~(c), before resuming with Fig.~\ref{fig:contractions}~(d). Nonetheless, this does not affect the total number of steps to perform decoding, each of which has some well-defined, fixed cost.
\end{IEEEproof}
\end{lemma}

Finally, an additional step allows us to see that decoding in both cases has computational cost that scales as $N\log_2 N$. This relies on storing the results of previous computations for later bits.
\begin{lemma}
A full sweep of successive cancellation decoding can be performed with computational cost $N\log_2 N$ for both the Polar and Convolutional Polar codes. 
\begin{IEEEproof}
For the Polar code, there are  only two possible calculations to perform at the lowest layer, each involving $N/2$ of one of the tensor network contractions in Fig.~\ref{fig:contractions}~(a). This is a total of $N$ calculations. The next layer has four possibilities, depending on which bit is undergoing successive cancellation decoding, but each requires only $N/4$ tensor contractions, making a total cost of $N$. At each layer the number of possibilities double while the number of CNOT gates halves, so the cost is constant. There are $\log_2 N$ layers so the total cost is proportional to $N \log_2 N$.

For the Convolutional Polar code, the step at the lowest layer need only be performed once, and involves $N/2$ contractions (Fig.~\ref{fig:contractions}~(b)). At the next layer, we have four possibilities: two bulk transformations that end with a 3-bit distribution (Fig.~\ref{fig:contractions}~(c)) and two boundary transformations that end with a 2-bit distribution (Fig.~\ref{fig:contractions}~(e)). In any of these cases, there are $N/4$ diagrams to compute --- making a total of $N$ unique tensor network contraction steps. Above this point, when beginning with a 2-bit distribution, there are just two further possibilities: either remain in the 2-bit distribution or ascend to a 3-bit distribution (as the choice of left or right is given fully determined by the choices in lower layers). When beginning with a 3-bit distribution, again we have two possibilities (left or right). In either case, the number of possibilities doubles for each layer ascended while the number of diagrams halves, so precisely $N$ unique diagrams need by determined for each of these $\log_2 N - 2$ layers. The total number of required contraction steps is $N \log_2 N - N/2$.
\end{IEEEproof}
\end{lemma}

\section{Channel polarization}
\label{sec:polarization}

The breakthrough achievement of Arikan was to prove, under a well-defined and efficient decoding scheme, that the logical channels corresponding to individual bits polarize exponentially into either ``perfect'' or ``useless'' channels in the limit of very large Polar codes --- and further that the ratio of good to bad channels tends to the (symmetric) capacity. Here we present similar proofs for the case of Convolutional Polar codes.  Furthermore we will see that in the limit of very large block length, the error exponent is provably better for Convolutional Polar codes than it is for Polar codes. 

Our proof studies the effect of each polarization step constituting  the Convolutional Polar code. In the case of Polar codes, these are simply CNOT gates combining two channels of same mutual information. For Convolutional Polar codes, each transformation additionally involves CNOT gates connecting the two channels to neighbouring blocks respectively, see  \fig{primitives}.  Nevertheless, only identical channels get combined in each step forming two new types of channels. Moreover, just like for Polar codes, assuming a binary-input memoryless symmetric channel (BMSC) under successive cancellation decoding, the channels that get combined at every stage are independent. We aim at tracking the capacity of the resulting channels through these combining process. 

Our proof has two important differences to the one used for Polar codes. 1) While each elementary step combines channels with equal mutual information, channels with different mutual information are combined within each step, something that usually does not happen when investigating Polar codes. For this reason, we employ results derived for Polar codes for non-stationary channels \cite{AT14}. 2) The combination steps are not all equivalent due to the presence of boundary channels. The effect of these boundary channels is nonetheless negligible because they appear at a very low density except during the last few steps of polarization, once the channels are already almost perfectly polarized.

\subsection{Preliminaries}\label{Pre}

Let $W:X\rightarrow Y$ be an arbitrary  BMSC with input alphabet $X=\{0,1\}$. Then we denote with $I(W)$ the symmetric capacity of $W$ given by the mutual information between the input and output distributions. Further we define the Bhattacharyya parameter of $W$ as $Z(W)= \sum_{y \in Y} \sqrt{W(y|0)W(y|1)}$. Both the symmetric capacity and the Bhattacharyya parameter take values in $[0,1]$. Finally, we also use the binary entropy $h_2(p):=-p\log_2{(p)} - (1-p)\log_2{(1-p)}$.


Using the standard notation of Polar codes, we define a notation for the possible channels obtained from the elementary combining step in Convolutional Polar codes.
We denote the two main channels by
\begin{align}
W^{\widehat{-}} = \langle \langle W,W\rangle^{-} , \langle W,W\rangle^{-} \rangle^{+} \\
W^{\widehat{+}} = \langle \langle W,W\rangle^{-} , \langle W,W\rangle^{+} \rangle^{+}.
\end{align}
This allows us to write a bulk channel after $n$ combining steps, in analogy to Polar codes, as $W^s$ with $s\in\{ \widehat{+} ,\widehat{-}\}^n$, with $n=\log_2{N}$. The polarization of boundary channels is different from bulk channels. Without going into the details of the boundary polarization procedure, we use the notation $s\in \{\widetilde{+},\widetilde{-}\}$ when considering all the polarization procedures, irrespectively of whether they include boundary steps.

For any channel combination the following is true \cite{RU08}:
\begin{equation}
I(\langle W_1, W_2 \rangle^{-}) + I(\langle W_1, W_2 \rangle^{+}) = I(W_1) + I(W_2).
\end{equation}
It follows that 
\begin{equation}
I(W^{\widehat{-}}) + I(W^{\widehat{+}}) = I(W^-) + I(W^+) = 2 I(W). \label{chain}
\end{equation}

One of the main tools in proofing that Convolutional Polar codes are capacity achieving is the so-called \textit{Mrs. Gerbers Lemma}, first published in \cite{WZ73}, and adapted to our problem  in \cite{AT14}. 
\begin{lemma}[Mrs. Gerbers Lemma]\label{MGL}
Take two channels $W_1$ and $W_2$ with mutual information $I(W_1) = 1 - h_2(p_1)$ and $I(W_2) = 1 - h_2(p_2)$ respectively. Then,   
\begin{equation}
I(\langle W_1, W_2 \rangle^{-}) \leq 1 - h_2(p_1 * p_2),
\end{equation}
with the convolution $x*y=x(1-y)+(1-x)y$.
\end{lemma}
Since $0\leq p_1,p_2 \leq \frac{1}{2}$ we have $p_1*p_2 \geq \max\{ p_1,p_2 \}$, with equality if and only if $p_1$ or $p_2$ are extremal. It follows
\begin{align}
I(\langle W_1,W_2\rangle^-) \leq \min\{I(W_1),I(W_2)\} \nonumber\\
\leq \max\{I(W_1),I(W_2)\} \nonumber\\
\leq I(\langle W_1,W_2\rangle^+). 
\end{align}
Here the first and last inequality are only equalities when the channels $W_1$ and $W_2$ are extremal. We can conclude that 
\begin{align}
I(&\langle W_1,W_2\rangle^+) - I(\langle W_1,W_2\rangle^-) \nonumber\\
&\geq \max\{I(W_1),I(W_2)\} - \min\{I(W_1),I(W_2)\} \nonumber \\
&\;+ \eta(a,b),
\end{align}
where the function $\eta(a,b)$ depends solely on $a = \min\{I(W_1),I(W_2)\}$ and $b = \max\{I(W_1),I(W_2)\}$. 
In the special case when both channels are the same, and $I(W)\in [a,b]$ this easily simplifies to
\begin{align}
I(W^+)-I(W^-)\geq \kappa(a,b), 
\end{align}
with $\eta(a,b)$ and $\kappa(a,b)$ being strictly positive when $0<a<b<1$. 
In \cite{AT14}, this special case is used to show that Polar codes polarize for stationary channels, while the general case is needed to prove polarization for non-stationary channels. Due to the more complicated structure of Convolutional Polar codes, we will use both versions here to prove polarization for stationary channels under Convolutional Polar codes. 

Finally, for a boundary polarization procedure, one of the gates is trivial so either the $\langle W,W\rangle^{-}$ or the $\langle W,W\rangle^{+}$ have to be substituted by $W$. The calculation then goes similar to the previous one. Note that the constant in these cases can in general be smaller, but nevertheless it is non-zero whenever the underlying channels are not already polarized. 
We will denote by $\widetilde\eta(a,b)$ and $\tilde \kappa(a,b)$ the minimum over all constants for the different types of channels.

\subsection{Convolutional Polar codes are capacity achieving}\label{main}

Our goal is now to prove that the fraction of \textit{good} channels goes to the mutual information for infinite uses of the elementary transform, while the fraction of \textit{mediocre} channels goes to zero, then the code is called polarizing. This is ensured by the following theorem. 
\begin{theorem}[Convolutional Polar codes polarize]\label{one}
For any BMSC $W$ and any $0<a<b<1$, Convolutional Polar codes achieve polarization to capacity in the sense that
\begin{align}
\lim_{n\rightarrow\infty} \frac{1}{2^n} \#\{ s\in\{\widetilde{+} ,\widetilde{-}\}^{n} : I(W^{s})\in [0,a)   \} &= 1 - I(W) \nonumber\\
\lim_{n\rightarrow\infty} \frac{1}{2^n} \#\{ s\in\{\widetilde{+} ,\widetilde{-}\}^{n} : I(W^{s})\in [a,b]   \} &= 0 \nonumber\\
\lim_{n\rightarrow\infty} \frac{1}{2^n} \#\{ s\in\{\widetilde{+} ,\widetilde{-}\}^{n} : I(W^{s})\in (b,1]   \} &= I(W), \nonumber
\end{align}
where $n$ denotes the number of polarization steps. 
\end{theorem}

To prove that capacity is indeed achieved, we also need to show that the block error probability can be kept arbitrarily small. To track the error rate through the polarization process we have to additionally give bounds on the Bhattacharyya parameter after one step of channel combining. In \cite{A09a} bounds on the  Bhattacharyya parameter after one Polar transform were given. Here we extend these bounds to the combination of different channels. The proof is analogue to the one in \cite{A09a}. 
\begin{proposition}
For two BMSC $W_1$ and $W_2$, 
\begin{align}
Z(\langle W_1,W_2\rangle^+) &= Z(W_1) Z(W_2) \label{Z1}\\
Z(\langle W_1,W_2\rangle^-) &\leq Z(W_1)+Z(W_2) - Z(W_1)Z(W_2)\\ 
& \leq  2\max{\{Z(W_1),Z(W_2)\}}. \label{Z2}
\end{align}
\begin{IEEEproof}
We are considering two binary symmetric channels $W_1(y|x)$ and $W_2(y|x)$.
 The channel polarization combines these two channels into a two-bit channel
$$
P(y_1,y_2|u_1,u_2) = W_1(y_1|u_1+u_2)W_2(y_2|u_2).
$$
We are interested in the Bhattacharyya parameters of the channels $W^- = P(u_1|y_1,y_2)$ and $W^+ = P(u_2|y_1,y_2,u_1)$. Let's start by computing the second one, for any value of $u_1$:
\begin{align}
Z(\langle W_1,W_2\rangle^+)
& = \sum_{y_1,y_2} \sqrt{P(y_1y_2|u_1,0)P(y_1y_2|u_1,1)} \\
& = \sum_{y_1,y_2} \sqrt{W_1(y_1|u_1)W_2(y_2|0)W_1(y_1|u_1+1)W_2(y_2|1)}\\
& = \sum_{y_1} \sqrt{W_1(y_1|u_1)W_1(y_1|u_1+1)} \sum_{y_2} \sqrt{W_2(y_2|0)W_2(y_2|1)} \\
& = Z(W_1)Z(W_2).
\end{align}
For the first one, we have
\begin{align}
Z(\langle W_1,W_2\rangle^-)
& = \sum_{y_1,y_2} \sqrt{\sum_{u_2} P(u_2)P(y_1y_2|0,u_2)\sum_{v_2} P(v_2)P(y_1,y_2|1,v_2)} \\
&= \sum_{y_1,y_2} \sqrt{\sum_{u_2,v_2}\frac 14 W_1(y_1|u_2)W_2(y_2|u_2)W_1(y_1|v_2+1)W_2(y_2|v_2)} \\
& =\frac 12 \sum_{y_1,y_2} \sqrt{ W_1(y_1|0)W_2(y_2|0)+W_1(y_1|1)W_2(y_2|1)}\times \sqrt{W_1(y_1|1)W_2(y_2|0) + W_1(y_1|0)W_2(y_2|1)} \\
&\leq \frac 12 \sum_{y_1,y_2} \left[ \sqrt{W_1(y_1|0)W_2(y_2|0)} + \sqrt{W_1(y_1|1)W_2(y_2|1)}\right] \times  \left[ \sqrt{W_1(y_1|1)W_2(y_2|0)} + \sqrt{W_1(y_1|0)W_2(y_2|1)}\right] \label{eq:inequality}\\
& -  \sum_{y_1,y_2} \sqrt{W_1(y_1|0)W_1(y_1|1)W_2(y_2|0)W_2(y_2|1)} \\
& =\frac 12 \sum_{y_1,y_2}\left[ W_2(y_2|0)\sqrt{W_1(y_1|0)W_1(y_1|1)} + W_2(y_2|1)\sqrt{W_1(y_1|0)W_1(y_1|1)} \right.\\
& \left. \quad\quad\quad\quad + W_1(y_1|0)\sqrt{W_2(y_2|0)W_2(y_2|1)} + W_1(y_1|1)\sqrt{W_2(y_2|0)W_2(y_2|1)} \right] -Z(W_1)Z(W_2)\\ 
& = Z(W_1)+Z(W_2) - Z(W_1)Z(W_2).
\end{align}
The inequality \eq{inequality} follows from the equality
\begin{align}
\left[\sqrt{(\alpha\beta+\delta\gamma)(\alpha\gamma + \delta\beta)}\right]^2 + 2\sqrt{\alpha\beta\gamma\delta}(\sqrt\alpha - \sqrt\delta)^2(\sqrt\beta - \sqrt\gamma)^2 
 = \left[(\sqrt{\alpha\beta} + \sqrt{\delta\gamma})(\sqrt{\alpha\gamma} + \sqrt{\delta\beta}) - 2\sqrt{\alpha\beta\gamma\delta}\right]^2.
\end{align}
We have also made used of the trivial equality $\frac 12 \sum_y [W(y|0) + W(y|1)] = \sum_y p(y) = 1$ for any symmetric channel $W$. 
\end{IEEEproof}
\end{proposition}
This now allows us to put bounds on combining steps for the Convolutional Polar code,
\begin{align}
Z(W^{\widehat{-}}) &\leq 2\max{\{Z(W^+),Z(W^-)\} } \nonumber\\ 
&= 2\max{\{Z(W)^2,2Z(W)\} } = 4Z(W) \label{eq:W-}
\end{align}
and
\begin{align}
Z(W^{\widehat{+}}) &= Z(W^+)Z(W^-) \nonumber\\
&\leq Z(W)^2 2Z(W) = 2 Z(W)^3. \label{eq:W+}
\end{align}
With this we can investigate the block error rate, resulting in the following. 
\begin{theorem}[Asymptotically vanishing block error rate for Convolutional Polar codes]\label{two}
Consider a BMSC $W$, then for any fixed $\beta'<\frac{1}{2}\log_2{(3)}$
\begin{align}
P_e(N,A,u_{A^c}) \leq O(2^{-N^{\beta'}}).
\end{align}
Here $N$ is the length of the code, $A$ is the set fixing the indices of the information bits and $u_{A^c}$ is the vector of frozen bits. 
\end{theorem}
Note that this converges in general faster than the rate for Polar codes, where $\beta<\frac{1}{2}$. 
An interesting problem would be to determine the exact error exponent and to compare it with the one achievable by Polar codes.

\subsection{Proof of Theorem \ref{one}}

We now proof that Convolutional Polar codes are capacity achieving. 
Therefore we essentially follow the technique described in \cite{AT14}.
Given a binary symmetric channel $W$ and $0<a<b<1$, define
\begin{align}
\alpha_n(a,b) &:=  \frac{1}{2^n} \#\{ s\in\{\widetilde{+} ,\widetilde{-}\}^{n} : I(W^{s})\in [0,a)   \}  \nonumber\\
\theta_n(a,b) &:=  \frac{1}{2^n} \#\{ s\in\{\widetilde{+} ,\widetilde{-}\}^{n} : I(W^{s})\in [a,b]   \}  \nonumber\\
\beta_n(a,b) &:=  \frac{1}{2^n} \#\{ s\in\{\widetilde{+} ,\widetilde{-}\}^{n} : I(W^{s})\in (b,1]   \} \nonumber
\end{align}
Additionally we will need the quantities 
\begin{align}
\mu_n = \frac{1}{2^n} \sum_{s\in\{ \widetilde{+} ,\widetilde{-}\}^n} I(W^s) \\
\nu_n = \frac{1}{2^n} \sum_{s\in\{ \widetilde{+} ,\widetilde{-}\}^n} I(W^s)^2
\end{align}
Immediately a simple calculation using Eq.~(\ref{chain}) yields
\begin{align}
\mu_{n+1} &= \frac{1}{2^{n+1}} \sum_{s\in\{ \widetilde{+} ,\widetilde{-}\}^{n+1}} I(W^s) \nonumber\\
&= \frac{1}{2^{n}} \sum_{t\in\{ \widetilde{+} ,\widetilde{-}\}^{n}} \frac{1}{2}(I(W^{t\widetilde{+}})+I(W^{t\widetilde{-}})) \nonumber\\
&= \frac{1}{2^{n}} \sum_{t\in\{ \widetilde{+} ,\widetilde{-}\}^{n}} I(W^t)  \nonumber\\
&= \mu_n
\end{align}
Further we define 
\begin{align}
\Delta W := \frac{1}{2}[ I(W^{\widetilde{+}}) - I(W^{\widetilde{-}}) ]
\end{align}
One of the crucial points of this technique is to find a lower bound on $\Delta W$. This can be done by invoking the Mrs. Gerbers Lemma (Lemma \ref{MGL}).  This is where we have to be careful that indeed each of the different combinations picks up a non-zero constant whenever the channels are not polarized. For the $\widehat{+}$ and $\widehat{-}$ and for $I(W)\in [a,b]$ this can be done as follows
\begin{align}
\Delta W 
&= \frac{1}{2}[ I(\langle \langle W,W\rangle^{-} , \langle W,W\rangle^{+} \rangle^{+}) - I(\langle \langle W,W\rangle^{+} , \langle W,W\rangle^{-} \rangle^{-}) ] \nonumber\\
&\geq \frac{1}{2} \left( \max(I(W^+),I(W^-)) - \min(I(W^+),I(W^-)) \right) + \eta(a,b) \nonumber\\
&= \frac{1}{2} \left( I(W^+) - I(W^-) \right)   + \eta(a,b)  \nonumber\\
&\geq \kappa(a,b)  + \eta(a,b)
\end{align}

A similar bound holds for boundary channels, but with the constant $\tilde \kappa(a,b)  + \tilde \eta(a,b)$. Let us denote the minimum of all such constants $\alpha(a,b)$. Now, this allows us to give a bound on $\nu_{n+1}$, 
\begin{align}
\nu_{n+1} &=  \frac{1}{2^{n+1}} \sum_{s\in\{ \widetilde{+} ,\widetilde{-}\}^{n+1}} I(W^s)^2 \nonumber\\
&=  \frac{1}{2^{n}} \sum_{t\in\{ \widetilde{+} ,\widetilde{-}\}^{n}} \frac{1}{2}[I(W^{t\widetilde{+}})^2 + I(W^{t\widetilde{-}})^2 ] \nonumber\\
&=  \frac{1}{2^{n}} \sum_{t\in\{ \widetilde{+} ,\widetilde{-}\}^{n}} [I(W^{t})]^2 + [ \Delta (W^{t})]^2 \nonumber\\
&\geq \nu_n + \theta_n(a,b)\alpha(a,b)^2 \label{LB}
\end{align}
Here the last equality follows from the identity $\frac{1}{2}(a^2+b^2)= (\frac{1}{2}(a+b))^2 +  (\frac{1}{2}(a-b))^2$.
Since the second term in Eq.~(\ref{LB}) is always non-negative (and even strictly positive whenever the underlying channels are not already extremal), we get $\nu_{n+1}\geq\nu_{n}$ for all $n$. Furthermore $\nu_{n}$ is upper-bounded by $1$, so the series $\nu_{n}$ converges. 

As in \cite{AT14} this allows us to state bounds on the set $\theta_n$,
\begin{align}
0 \leq \theta_n(a,b) \leq \frac{\nu_{n+1} - \nu_n}{\alpha(a,b)^2}.
\end{align}
Since the $\nu_n$ converge for $n$ going to infinity we can state
\begin{align}
\lim_{n\rightarrow\infty} \theta_n(a,b) = 0.
\end{align}
Additionally an upper bound on the mutual information of the channel can be given by
\begin{align}
I(W) = \mu_n &\leq a\alpha_n(a) + b\theta_n(a,b) + 1\beta_n(b) \nonumber\\
&= a + (b-a)\theta_n(a,b) + (1-a)\beta_n(a,b).
\end{align}
Since we know the asymptotic behaviour of $\theta_n$ and we can choose $a$ arbitrary small, it follows that
\begin{align}
\lim_{n\rightarrow\infty} \beta_n(b) \geq I(W). 
\end{align}
Similarly, we get
\begin{align}
1 - I(W) &= 1 - \mu_n \nonumber\\
&\leq \alpha_n(a) + (1-a)\theta_n(a,b) + (1-b) \beta_n(b) \nonumber\\
&= b\alpha_n(a) + (b-a)\theta_n(a,b)  + (1-b). 
\end{align}
Again using the properties of $\theta_n$ and $b$ we can see that
\begin{align}
\lim_{n\rightarrow\infty} \alpha_n(a) \geq 1 - I(W).
\end{align}
Finally, since $\alpha_n(a) + \beta_n(b) \leq 1$ we can conclude that 
\begin{align}
\lim_{n\rightarrow\infty} \alpha_n(a) &= 1 - I(W) \\
\lim_{n\rightarrow\infty} \beta_n(b) &= I(W). 
\end{align}
This concludes the proof of polarization. 

\subsection{Proof of Theorem \ref{two}}

To show that the block error probability vanishes asymptotically we will combine two results from the investigation of Polar codes. 
Since Convolutional Polar codes use the same successive cancellation decoder as Polar codes, we can directly use a result from Arikan's original polar coding work~\cite{A09a}. Proposition 2 in~\cite{A09a} states that for any  discrete BMSC $W$, code length $N$ and $A$ being the set of synthesized channels used for transmitting information, the block error probability under successive cancellation decoding is bounded by
\begin{align}
P_e(N,A)\leq\sum_{i\in A}Z(W_N^{(i)}),
\end{align}
which implies that there exists an assignment for the choice of frozen indices $u_{A^c}$ such that 
\begin{align}\label{A}
P_e(N,A,u_{A^c})\leq\sum_{i\in A}Z(W_N^{(i)}).
\end{align}
The proof of the Proposition only relies on the structure of successive cancellation decoding and therefore also applies to our setting. 

The second part of the proof is now concerned with bounding $Z(W_N^{(i)})$. For Polar codes this has been investigated originally in \cite{A09a} and more precise in \cite{AT09} and \cite{GX15}. We choose the approach used in \cite{AT09} and \cite{GX15}, since better bounds are achievable this way. The proof consists of two steps, the first one called \textit{rough polarization} which is based on tracking the channel entropy through the combinations. As the second step we will prove \textit{fine polarization} by investigating the Battacharyya parameter. 
 
For rough and fine polarization we will mostly consider the bulk coding steps. As we saw at \eq{fraction}, only an exponentially vanishing fraction of the channels are boundary channels during the first few polarization steps : boundary channels only become relevant during the last few polarization steps, when the channels are already almost perfectly polarized. To prove rough polarization for bulk channels,  we will apply Proposition 5 from \cite{GX15}, which we can state as follows. 

\begin{proposition}[from \cite{GX15}]For a constant $\Lambda<1$, for all $\rho\in (\Lambda,1)$ there exists a constant $c_{\rho}$ such that for all binary symmetric channels $W$, all $\epsilon>0$ and $n\geq c_{\rho} \log_2{(\frac{1}{\epsilon})}$ then 
\begin{align}\label{rough0}
\Pr_i\left( Z(W_n^{(i)}) \leq 2\rho^n\right) \geq I(W) - \epsilon. 
\end{align}
\label{prop:GX15}
\end{proposition}

To apply this Proposition to our setting we simply have to make sure that our coding scheme satisfies the following conditions (serving as analogue to Lemma 6 from \cite{GX15} for the channels $W^{\widehat{+}}$ and $W^{\widehat{-}}$, when proving rough polarization), 
\begin{align}
H(W^{\widehat{+}}) \leq H(W) - \alpha(W) \\
H(W^{\widehat{-}}) \geq H(W) + \alpha(W), 
\end{align}
where $\alpha(W)$ is a channel dependent constant, derived by using the Mrs. Gerbers Lemma and the conclusions drawn from it in Section \ref{Pre}. Here $H(W^{\widehat{+}})$ is the conditional entropy of the channel $W^{\widehat{+}}$, which for a symmetric channel is simply given by $1-I(W^{\widehat{+}})$.
Note the following
\begin{align}\label{pl}
H(W^{\widehat{+}}) &\leq \min{(H(W^+),H(W^-))} = H(W^+) \nonumber\\
&\leq H(W) - \alpha(W), \\
H(W^{\widehat{-}}) &\geq \max{(H(W^+),H(W^-))} = H(W^-) \nonumber\\
&\geq H(W) + \alpha(W). \label{mi}
\end{align} 
The last inequalities in Eqs. (\ref{pl}) and (\ref{mi}) follow from Lemma 6 in \cite{GX15}.
This is sufficient to apply the results from \cite{GX15}. Nevertheless note that the inequalities Eqs.~(\ref{pl}) and (\ref{mi}) are very rough and a more careful calculation could lead to a better error bound.

We additionally need to modify the original lemma a bit since we are planning to apply it only to bulk channels. Therefore we assume a \textit{worst-case} scenario, in which all boundary channels have high Bhattacharyya parameter (this is in practice very unlikely, since the boundary channels are symmetrically distributed to both sides of the code). This assumption has the following impact on Proposition \ref{prop:GX15}:
\begin{align}\label{rough}
\Pr_{i\in {\rm bulk}}\left( Z(W_n^{(i)}) \leq 2\rho^n\right) \geq I(W) - \epsilon - f(n), 
\end{align}
with $\lim_{n\rightarrow \infty} f(n) = 0$ whenever $m$ grows strictly slower than linear in $n$. (E.g. later we will look at $m = n^{\frac{3}{4}}$, therefore $f(n)=2^{n^{\frac{3}{4}}+1-n}$, based on \eq{fraction}.) This proves \textit{rough} polarization for all bulk channels. \\

In the next step we will  prove \textit{fine} polarization for Convolutional Polar codes,  along the lines of \cite{AT09} and \cite{GX15}. We start with Eqs.~(\ref{eq:W-}) and (\ref{eq:W+}). 
Note that compared to the Polar code case (see e.g. \cite{AT09}) we pick up an additional factor $2$ in each step, but, on the other hand, each $\widehat{+}$-channel has the original Bhattacharyya parameter cubed instead of squared. This will imply a faster polarization rate.

As in \cite{AT09}, we will divide the $n-m$ last polarizing steps into intervals of $\sqrt n$ steps and analyze the effect of each interval.  Thus, fix $\beta<\frac{1}{2}$, $a_n = \sqrt{n}$, $k = (n-m)/a_n$ and $k$ intervals $J_j = \{ m+(j-1)a_n, \dots, m+ja_n-1\}$. With $E_j$ we denote the event that $\#_{i\in J_j}\{s_i = \widehat{+}\} < a_n\cdot\beta$. In the absence of boundary gates, we get from a Chernoff-Hoeffding-type argument,
\begin{align}
P(E_j)\leq 2^{-a_n[1-h_2(\beta )]}.
\end{align}
Further we define the event $G_1:=\cap_j E_j^c$ and observe that it has at least probability $1-k2^{-a_n[1-h_2(\beta)]}$. 

Again we want to focus on the elementary bulk coding steps. We therefore consider all except the last polarization interval. From the previous definitions and \eq{fraction} we can see that after each interval $j$, the fraction of bulk channels is
\begin{equation}
1 - 2^{m + j a_n -n}.
\end{equation}
In particular, before the last polarization interval, a fraction $1-2^{-\sqrt{n}}$ of the channels are in the bulk. Let us call this set of bulk channels   $G_2$. Furthermore we denote $G=G_1\cap G_2$, which contains almost all channels when $n$ is large enough. In the following we will focus on channels in the set $G$. 

During each interval $J_j$, for a sequence $s\in G$, the value $Z(W^s)$ is derived by applying $s_i = \widehat{+}$ at least $a_n\beta$ times and $s_i = \widehat{-}$ at most $a_n(1-\beta)$ times. We will denote $W^s$ for any $s\in G$ by $W^G$, and use a subscript to denote the number of polarization steps. Hence, 
\begin{align}
\log_2{Z(W^G_{m+(j+1)a_n})} \leq 3^{a_n\beta}[\log_2{2^{a_n} Z(W^G_{m+ja_n})} +a_n(1-\beta)], \nonumber
\end{align}
where the additional factor $2^{a_n}$ inside the logarithm and the higher base in front distinguish the inequality from the Polar coding case, it reflects precisely the factor $2$ in each polarization step and the higher exponent for $s_i = \widehat{+}$. Now, following the calculation in \cite{AT09}, only taking into account transformations taking place in the first $k-1$ intervals, we derive
\begin{align}
\log_2{Z(W^G_{n})} &\leq 3^{(n-m-\sqrt{n})\beta}[\log_2{2^{a_n}Z(W_m)} + a_n(1-\beta)\sum_{j=1}^k 3^{ja_n\beta}] \nonumber\\
&\leq 3^{(n-m-\sqrt{n})\beta}[\log_2{2^{a_n}Z(W_m)} + a_n ] \nonumber\\
&= 3^{(n-m-\sqrt{n})\beta}[\log_2{Z(W_m)} + 2a_n ].
\end{align}
Finally, fix $m = n^{\frac{3}{4}}$, $\rho = \frac{7}{8}$ and define $\tilde G = \{ Z(W_m)\leq (\frac{7}{8})^m\} \cap G$, then for large $n$ we have $\log_2{Z(W^{\tilde G}_m)} \leq -n^{\frac{3}{4}}\log_2{(\frac{7}{8})}$. To justify the application of Proposition 5 from \cite{GX15}, note that all boundary channels after the first $m$ coding steps, who might have polarized slower, are not in the set $\tilde G$, since they are a strict subset of the channels already excluded from $G$ by definition.  
Therefore, for sufficiently large $n$, 
\begin{align}
\log_2{Z(W_{n})} &\leq 3^{(n-m-\sqrt{n})\beta}[-n^{\frac{3}{4}} \log_2{(\frac{7}{8})} + 2n^{\frac{1}{2}}] \nonumber\\
&\leq -3^{n\beta} o(1). 
\end{align}
Since the probability of $G$ converges to $1$, the probability of $\tilde G$ approaches $1-\epsilon$. It follows that 
\begin{align}
Z(W_n) &\leq 2^{-2^{\beta n\log_2{(3)}}} \nonumber\\
&= 2^{-2^{\beta' n}}
\end{align}
for any fixed $\beta' < \frac{1}{2}\log_2{(3)}$. 
Noting that $n=\log_2{N}$ we can use the above result to upper bound the block error rate of Convolutional Polar codes as follows
\begin{align}
P_e(N,A,u_{A^c}) &\leq\sum_{i\in A}Z(W_N^{(i)}) \nonumber\\
& \leq O(2^{-N^\beta}),
\end{align}
for some $\beta < \frac{1}{2}\log_2{(3)}$. This concludes the proof.

\section{Numerical simulation of error-correction performance}
\label{sec:results}

In this Section, we  present numerical results contrasting the finite-size performance of Polar and Convolutional Polar codes. Two type of results are presented. The first type provides exact numerical bounds on the error probability of the code under an erasure channel. The second type is obtained from Monte Carlo sampling and are applicable to any BMSC.  In all cases, we observe notable improvements using convolution.

\subsection{Exact numerical calculations for erasure channel}
\label{sec:erasure}

Here we study the channel polarization for both the Polar and Convolutional Polar codes using the erasure channel. We observe that channel polarization is stronger under the Convolutional Polar code resulting in reduction in the expected error rate.   

In~\cite{A09a}, it is shown that the CNOT gate used in a polarization step transforms two erasure channels with erasure rates $\epsilon_L$ and $\epsilon_R$ (on the left and right, respectively)  into two new effective erasure channels with erasure rates $\epsilon_L^{\prime} = \epsilon_L \epsilon_R$ and $\epsilon_R^{\prime} = \epsilon_L + \epsilon_R - \epsilon_L \epsilon_R$. The transformation is slightly more complicated for the Convolutional Polar, but can nonetheless be performed efficiently. 

Under the erasure channel, the value of a bit is either known or not. We will need to consider slightly more complex situations, for instance we might know the sum of two-bit values while not knowing either. In general, our state of knowledge can be summarized by linear constraints. In the above example, we would have $(1,1)\cdot (x_1,x_2) = a$. In general, for a collection of $n$ bits, our state of knowledge can be written $Cx = a$ where $C$ is a $k\times n$ matrix with $k\leq n$ and $a$ is a $k$-bit vector. Note that the actual value of the vector $a$ is not important in our analysis since we only care about whether a quantity is known or not, and not about its actual value.

Consider the convolved polar transforms illustrated on \fig{contractions} (d). They takes as input two 3-bit probability distributions $p_a$ and $p_b$, which can be combined into a 6-bit distribution $q = p_a\otimes p_b$, and performs a linear invertible transformation $M$ from 3 CNOT gates. Suppose that prior to the application of $M$, the state of knowledge of the 6 input bit was described by the equation $Cx = a$. Then, the state of knowledge of the output $y = Mx$ is described by $C'y = a$ where $C' = CM^{-1}$.  Focus now on the transformation on the left of  \fig{contractions} (d). After the linear transformation, the first two bits are fixed to known valuers $x_1$ and $x_2$ while the last bit is ignored. To understand the effect of this on our state of knowledge, we can put the matrix $C'$ in a standard form using row manipulations
\begin{equation}
C' \sim \left(
\begin{array}{c|c|c}
A_{(k-u)\times 2}&B_{(k-u)\times 3}&{\bf 0}_{(k-u)\times 1}\\
\hline
D_{u\times 2}&E_{u\times 3} & 1_u\\
\end{array}
\right),
\end{equation}
where the subscript indicates the dimension of the matrix size,  $u=0$ or $1$, and $1_u$ denotes the $u\times u$ identity matrix. Then, the matrix $B$ represents our new state of knowledge for bits $y_3$, $y_4$ and $y_5$. A similar reasoning applies to the transformation on the right of \fig{contractions} (d), the difference being that only the first bit is fixed and the last two bits are ignored. In that case, $u$ can take the value 0, 1 or 2, and the dimensions are adjusted to $A_{(k-u)\times 1}$ and $D_{u\times 1}$.

There are 16 matrices $C$ of dimensions $k\times 3$ for $k\leq 3$ which are distinct under row manipulations, see the Appendix. Our study of the erasure channel is based on assigning probabilities to these 16 possible states of knowledge, and evolving these distributions through convolutional polar transforms as described above. Initially, each bit is erased with probability $p$, so a collection of 3 bits has the following distribution of states of knowledge:
\begin{align}
p_1 = p^3, \quad&B_1 = \emptyset, \\
p_2 = (1-p)p^2, \quad &B_2 = \left(\begin{array}{ccc} 1&0&0 \end{array}\right), \\
p_3 = (1-p)p^2, \quad &B_3 = \left(\begin{array}{ccc} 0&1&0 \end{array}\right), \\
p_4 = (1-p)p^2, \quad &B_4 = \left(\begin{array}{ccc} 0&0&1 \end{array}\right), \\
p_9 = (1-p)^2p, \quad &B_9 = \left(\begin{array}{ccc} 1&0&0\\0&1&0 \end{array}\right), \\
p_{10} = (1-p)^2p, \quad &B_{10} = \left(\begin{array}{ccc} 1&0&0\\0&0&1 \end{array}\right), \\
p_{11} = (1-p)^2p, \quad &B_{11} = \left(\begin{array}{ccc} 0&1&0\\0&0&1 \end{array}\right), \\
p_{15} = (1-p)^3, \quad &B_{16} = \left(\begin{array}{ccc} 1&0&0\\0&1&0\\0&0&1 \end{array}\right) ,
\end{align}
with all 9 other $p_j=0$. Our algorithm proceeds by combining pairs of these states of knowledge $B_a$ and $B_b$ into a 6-bit state of knowledge $C = B_a\oplus B_b$ with probability $p_ap_b$, and applying either the procedure corresponding to the right or left of \fig{contractions} (d) to produce a new state of knowledge $B$. The resulting transformations on the states of knowledge are shown in the Appendix. Similar, simpler procedures can be derived for the other combinations shown on \fig{contractions}. Using this procedure, given as input an independent erasure probability $p$, we are able to compute exactly the probability $p_j$ that output bit $x_j$ is erased conditioned on the fact that none of the bits $x_i$ $i<j$ were erased, i.e., the probability that bit $j$ is the first to be erased under successive cancellation decoding. 

\begin{figure}[t]
\centering
\includegraphics[width=0.5\doublecolumnwidth]{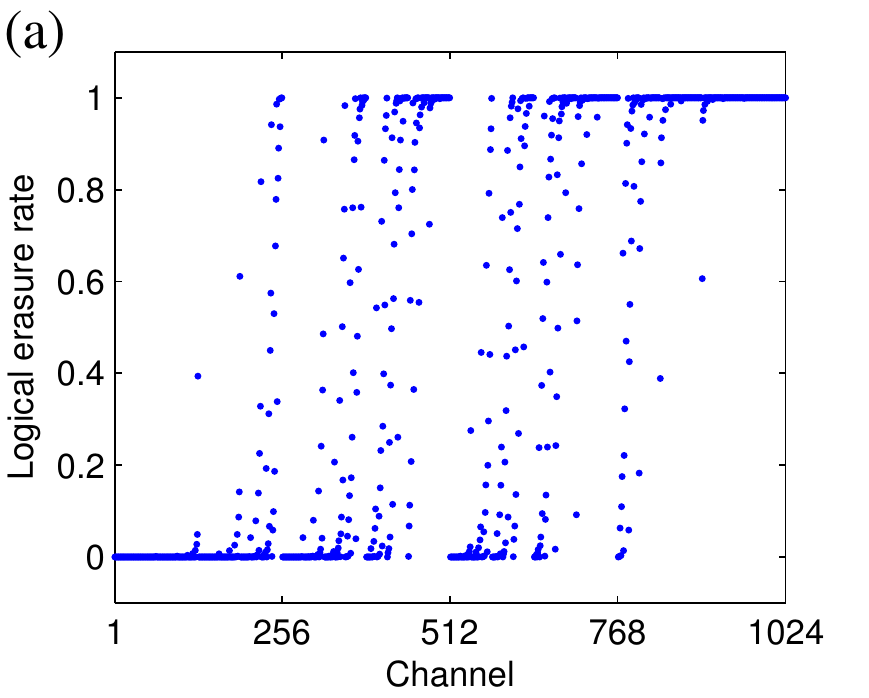}\includegraphics[width=0.5\doublecolumnwidth]{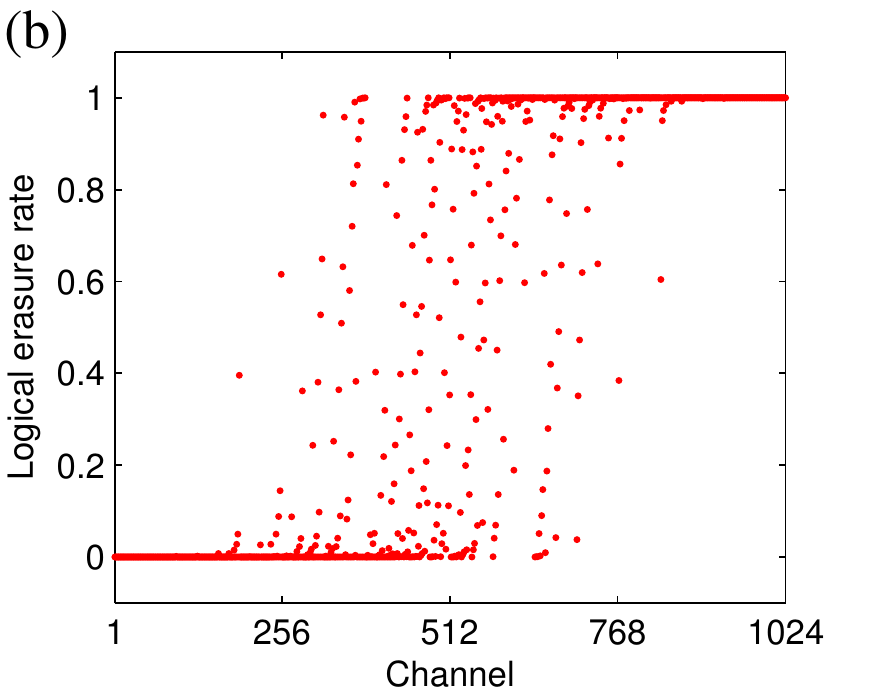}
\includegraphics[width=0.5\doublecolumnwidth]{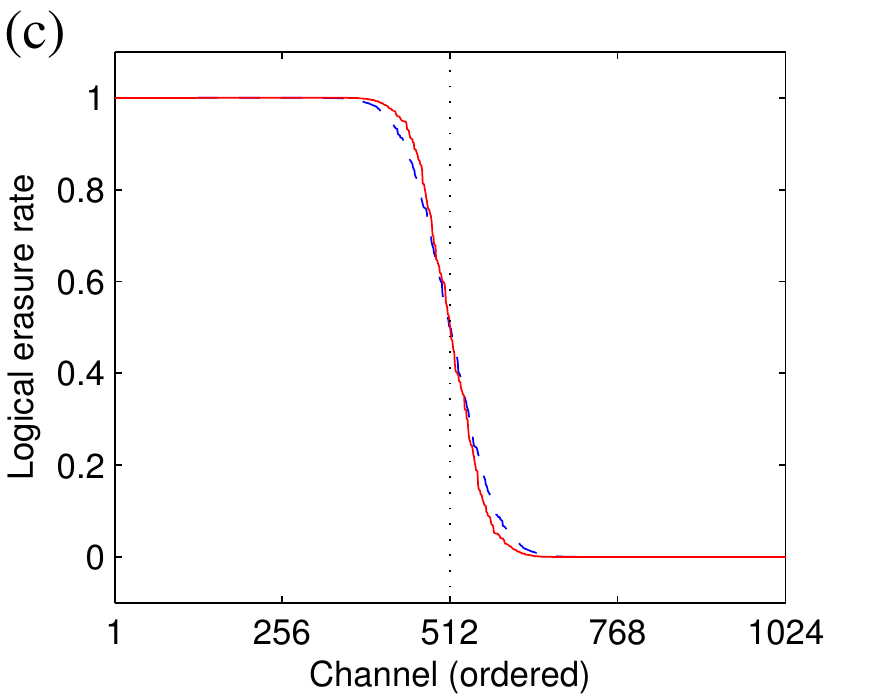}\includegraphics[width=0.5\doublecolumnwidth]{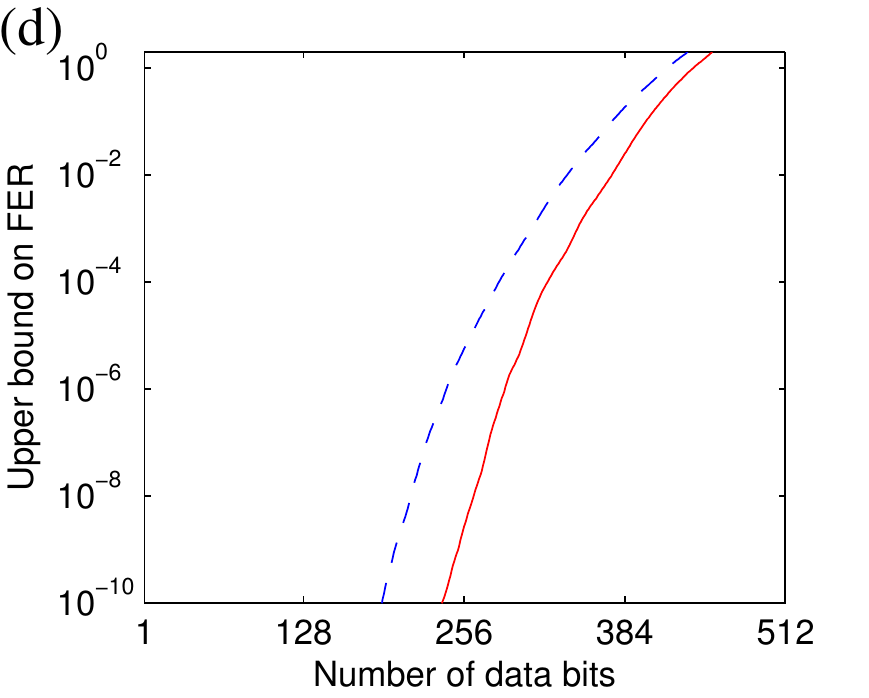}
\caption{Polarization of the logical channels under 1024-bit Polar and Convolutional Polar codes with the 50\% erasure channel. (a,b) The accuracy of each channel, parameterized by effective erasure rate of the logical channels under successive cancellation decoding, for (a) the Polar code and (b) the Convolutional Polar code. (c) The same data is plotted in descending order for the Polar (dashed, blue line) and Convolutional Polar (solid, red line) codes. The dotted vertical line corresponds to the capacity at 50\%. (d) The cumulative sum gives a simple upper bound to the frame-error rate (FER) for the specified number of data bits. \label{fig:channels} }
\end{figure}

We have used this method to study channel polarization of the erasure channel under Polar and Convolutional Polar coding. In \fig{channels}, we present results for the 50\% erasure channel over 1024 bits. In \fig{channels}~(a,b,c) we observe that the Convolutional Polar code contains somewhat fewer channels in the intermediate area between the perfect and useless limits, and further, that the good channels are a little more strongly localized on the left (and conversely the bad channels are localized on the right). This latter fact is particularly significant for successive cancellation decoding because it means more information regarding the frozen bits is available to the decoder when determining the data bits, reducing the gap in performance between maximum likelihood and successive cancellation decoding.

From these results we can also deduct a simple upper bound on the probability of at least one error in the block, or frame-error rate (FER).  By simply summing the probability of erasures over the $k$ data channels and noting that the chance of an error is \emph{at most} the sum of probabilities that a given data bit is the first to be decoded incorrectly, we arrive at an (over)-estimate of the FER. In \fig{channels}~(d) we show this sum for a range of data rates. Similarly, the maximum probability that a non-frozen bit is the  first to be decoded incorrectly provides a lower bound on the FER. While we show only lower bounds here, it is indistinguishable from the upper bound in most graphs shown here. Thus, these bounds suggest that the Convolutional Polar code can deliver a significant increase in the amount of data sent for a fixed error rate (especially for small target error rates).

\begin{figure}[t]
\centering
\includegraphics[width=0.5\doublecolumnwidth]{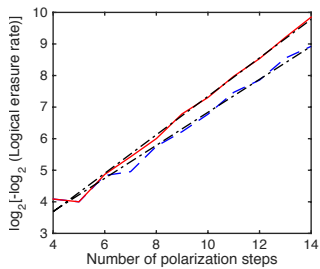}
\caption{Assuming that the FER scales as $P_e = 2^{-\gamma N^\beta}$ where the number of bits $N=2^n$ depends on the number of polarization steps $n$, we get $\log_2[-\log_2(P_e)] = \log_2\gamma + \beta n$. Using an upper bound for $P_e$, this graph plots $\log_2[-\log_2(P_e)] $ as a function of $n$ for (blue dash) Polar codes and (red full) Convolutional Polar codes. Also shown (black, dash-dot) are linear fits using $\gamma \approx 3.04$ and $\beta\approx 0.52$ for the Polar codes, and $\gamma\approx 2.37$ and $\beta\approx 0.61$ for Convolutional Polar codes. The channel is 50\% erasure and the encoding rate is $\frac 1{16}$.  \label{fig:error_exponent} }
\end{figure}

Finally, \fig{error_exponent} uses the upper bound (lower bound is indistinguishable) derived above to estimate the error exponent of the Polar and Convolutional Polar codes. As shown in \sec{polarization}, for a fixed encoding rate $k/n$ (less than the capacity) both codes have asymptotic frame-error rates scaling as $2^{-N^\beta}$ with $\beta\leq \frac 12$ for Polar codes and $\beta\leq \frac 12\log_23\approx 0.79$ for Convolutional Polar codes. While these scalings are asymptotic, the finite size scaling observed on \fig{error_exponent} reveals error exponents $\beta \approx 0.52$ for Polar codes and $\beta \approx 0.61$ for Convolutional Polar codes, providing good evidence that the enhance polarization is significant even at for relatively small block sizes. 

\subsection{Monte Carlo simulations}

Here we numerically compare the performance of the Polar and Convolutional Polar codes at protecting data from a variety of channels, focussing on finite-code length effects on codes between 256 and 8192 bits. For all our simulations we have used a simplified channel selection scheme that is independent of the details of the error model. Our scheme uses the symmetric bit flip channel with probability $p$ and evaluates, for each position $j$, the probability that bit $x_j$ is the first to be decoded incorrectly. This works by using an all-zero input and an output where the decoder believes each bit has an independent probability $p$ of being a 1, and $1-p$ of being a 0. For each bit $x_j$ we contract the corresponding tensor-network diagram, with $x_i = 0$ for $i<j$ and $x_i$ random for $i>j$. A more accurate estimate of the logical channel error rate for both the Polar code and Convolutional Polar code could be obtained through Monte Carlo sampling~\cite{A09a}, i.e. by sampling over the possible bit values $x_i$ with $i<0$ instead of fixing them to 0. Alternatively, more sophisticated techniques~\cite{Tal2013} could also be used for Convolutional Polar codes. However, we have found that this simplified procedure gives adequate results over all the channels we have investigated (for instance, performing better for the bit-flip channel than the data presented in \fig{channels}, which derives from the erasure channel). A slight improvement in performance can be observed by using the channel selection tailored to the specific error model in question, but the comparative performance between the Polar and Convolutional Polar codes remains very similar, so this procedure is adequate for the purposes of comparison.

\begin{figure}[t]
\centering
\includegraphics[width=0.5\doublecolumnwidth]{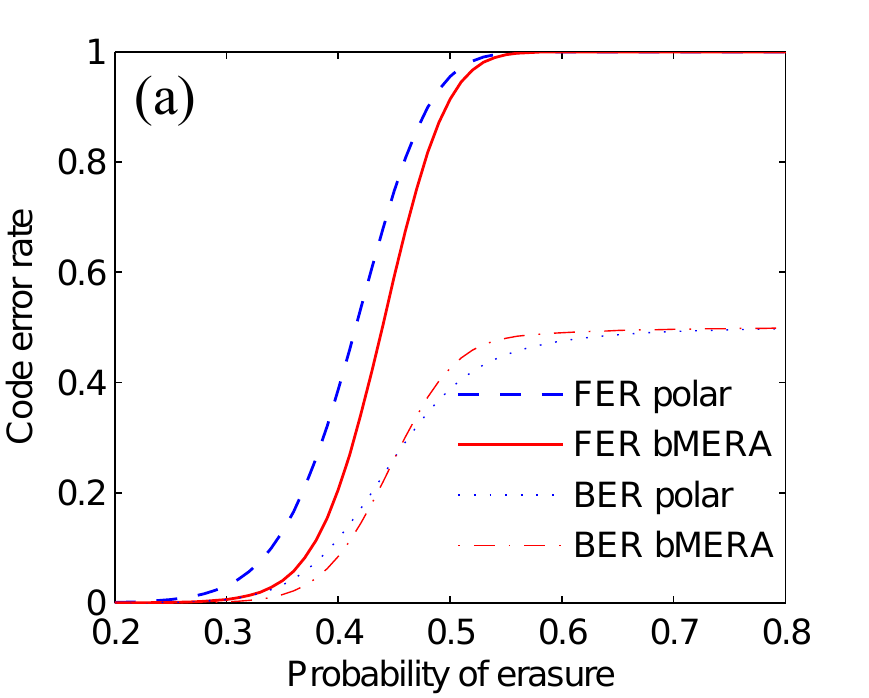}\includegraphics[width=0.5\doublecolumnwidth]{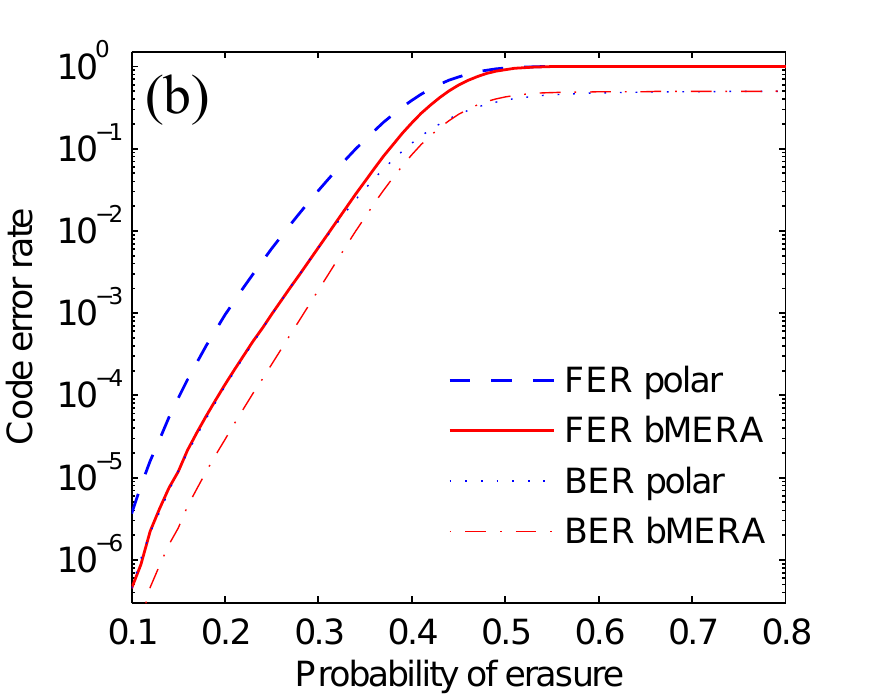}
\includegraphics[width=0.5\doublecolumnwidth]{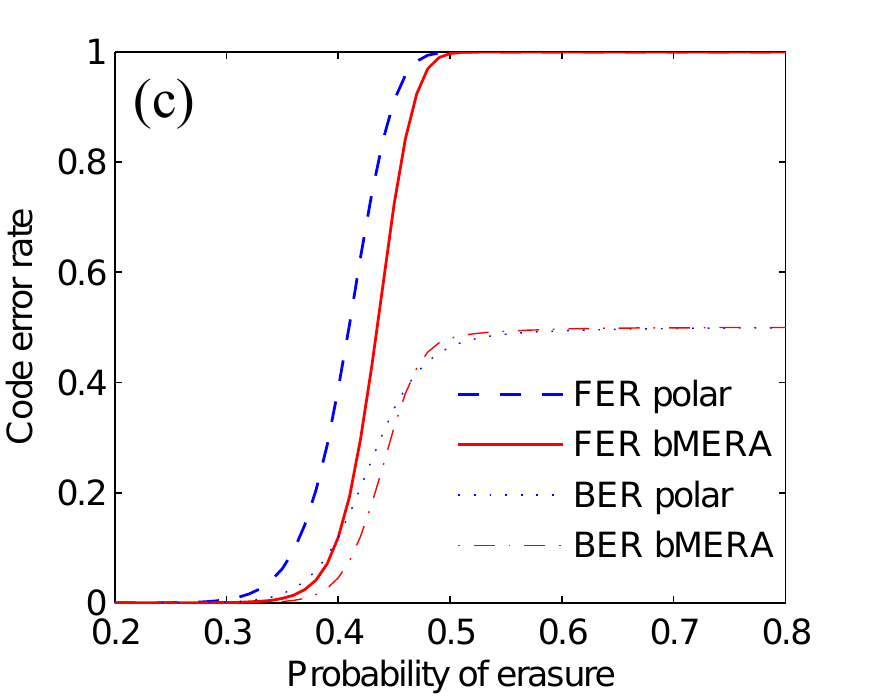}\includegraphics[width=0.5\doublecolumnwidth]{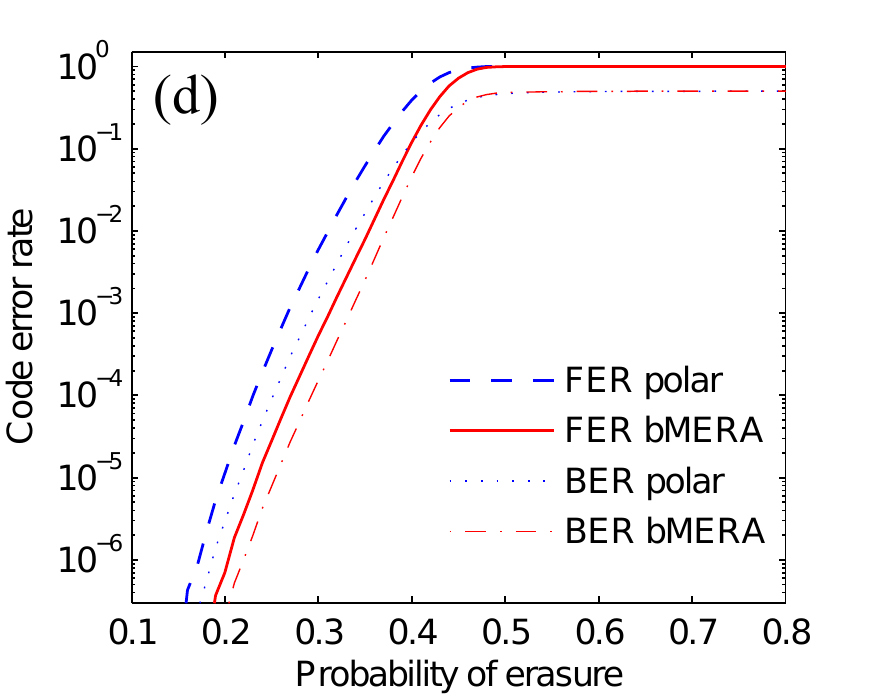}
\includegraphics[width=0.5\doublecolumnwidth]{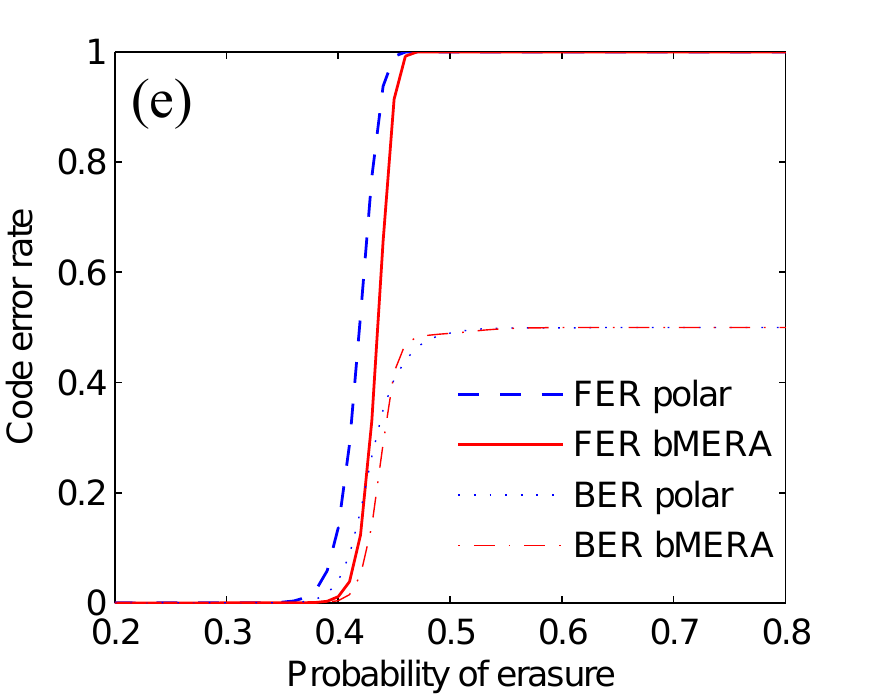}\includegraphics[width=0.5\doublecolumnwidth]{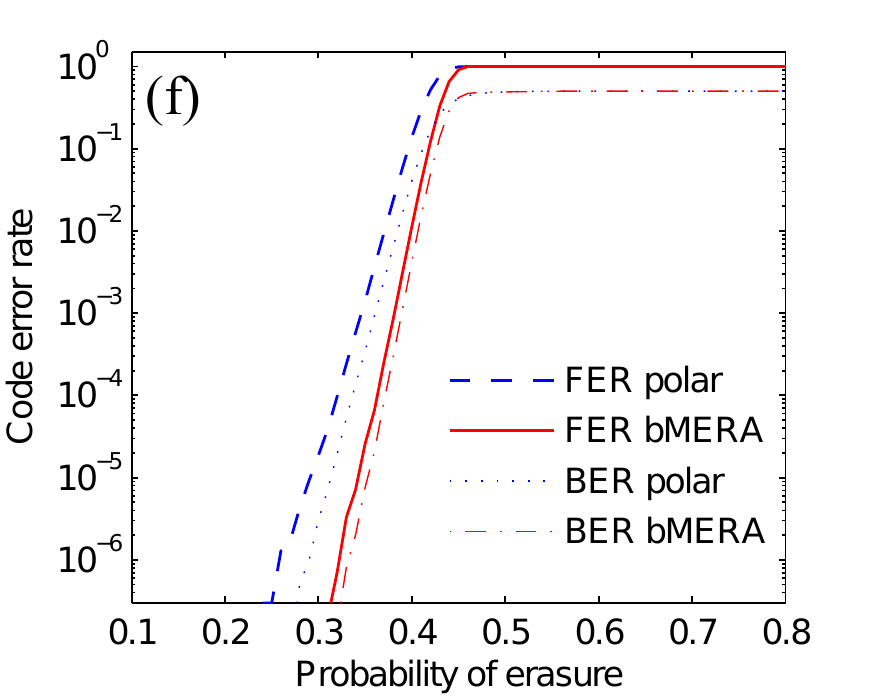}
\caption{Comparison of the performance of rate 1/2 Polar and Convolutional Polar codes of various sizes for the binary erasure channel. The encoded message contains (a,b) 256 bits. (c,d) 1024 bits and (e,f) 8192 bits. The capacity with erasure probability 0.5 corresponds to the code rate 1/2. \label{fig:classical_erasure} }
\end{figure}

The results for the binary erasure channel with code rate 1/2 are given in \fig{classical_erasure}. In all cases we observe several things. Finite-size effects are significant in both codes, with the waterfall region separating ``perfect" and ``useless" behaviour being somewhat below the capacity of the erasure channel (which suggests that erasure rates of up to 0.5 are tolerable for our encoding rate). Nonetheless, the threshold of the Convolutional Polar code is significantly closer to this value than the Polar code. On a logarithmic scale, it is evident that the performance in the low-error region is significantly better --- note the slope in \fig{classical_erasure}~(f) is significantly greater for the Convolutional Polar code. Neither code displays any evidence of an error floor (nor is it expected). Finally, both codes display a tendency for any error to be catastrophic --- involving errors on many bits. Indeed, the ratio between the bit error rate (BER) and frame error rate (FER) is very large for the Polar code and even higher (close to 0.5) for the Convolutional Polar code. This corresponds to either a perfectly decoded message or a completely scrambled one. Interestingly, this is the behaviour expected of a ``perfect" random code as Shannon envisaged, where the most likely messages are completely uncorrelated. 

\begin{figure}[t]
\centering
\includegraphics[width=0.5\doublecolumnwidth]{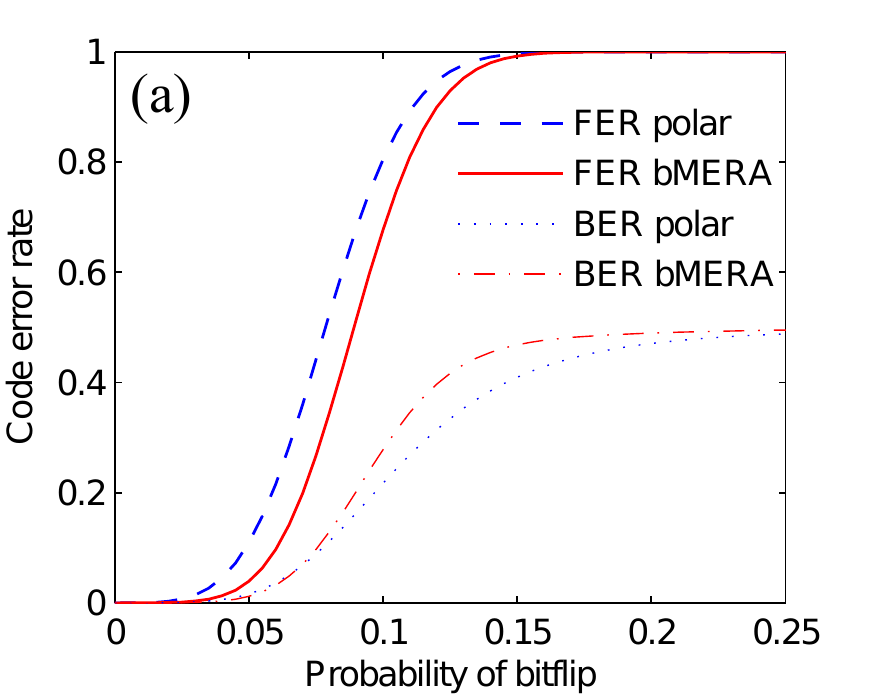}\includegraphics[width=0.5\doublecolumnwidth]{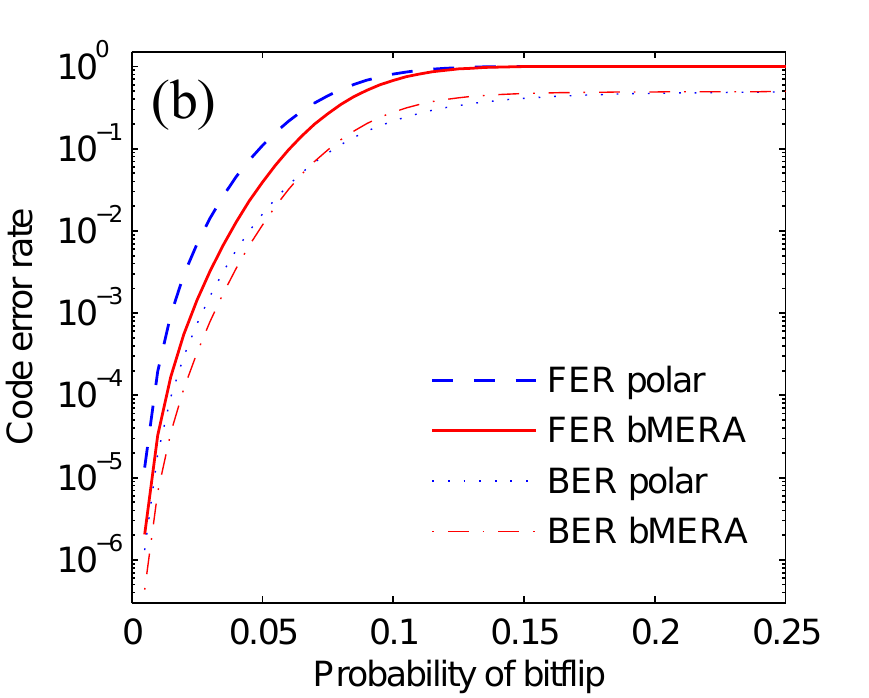}
\includegraphics[width=0.5\doublecolumnwidth]{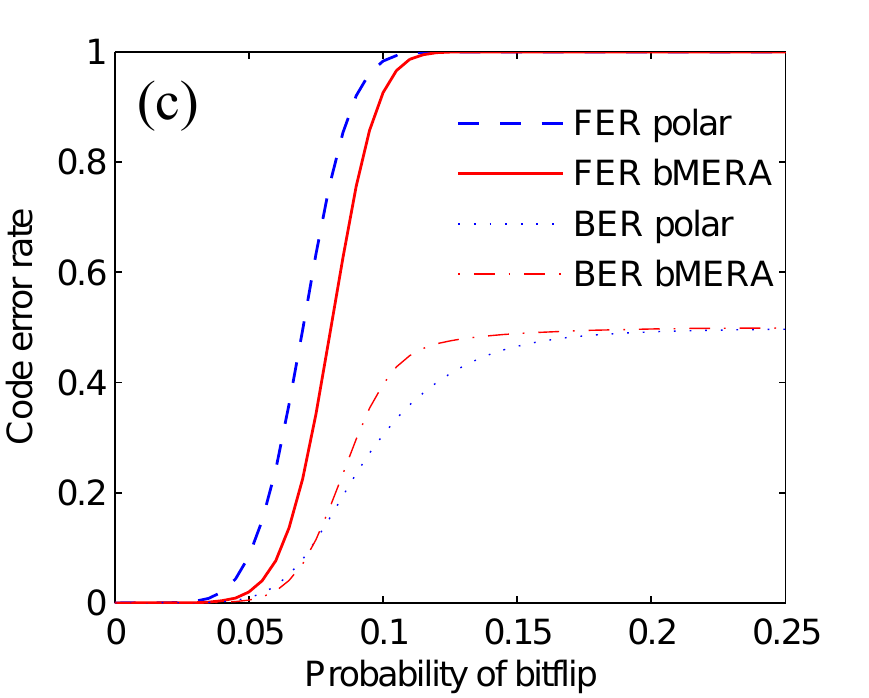}\includegraphics[width=0.5\doublecolumnwidth]{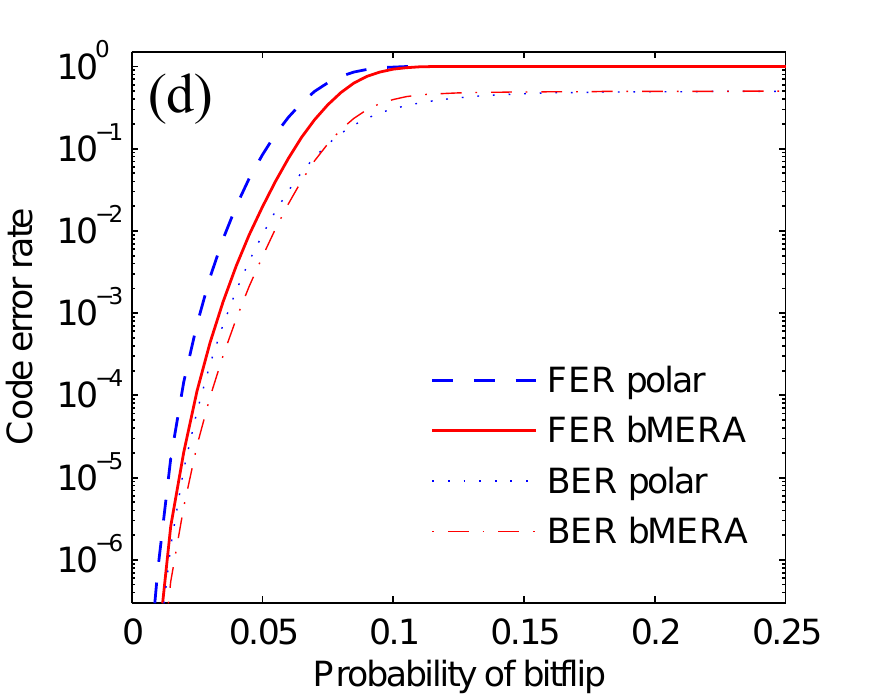}
\includegraphics[width=0.5\doublecolumnwidth]{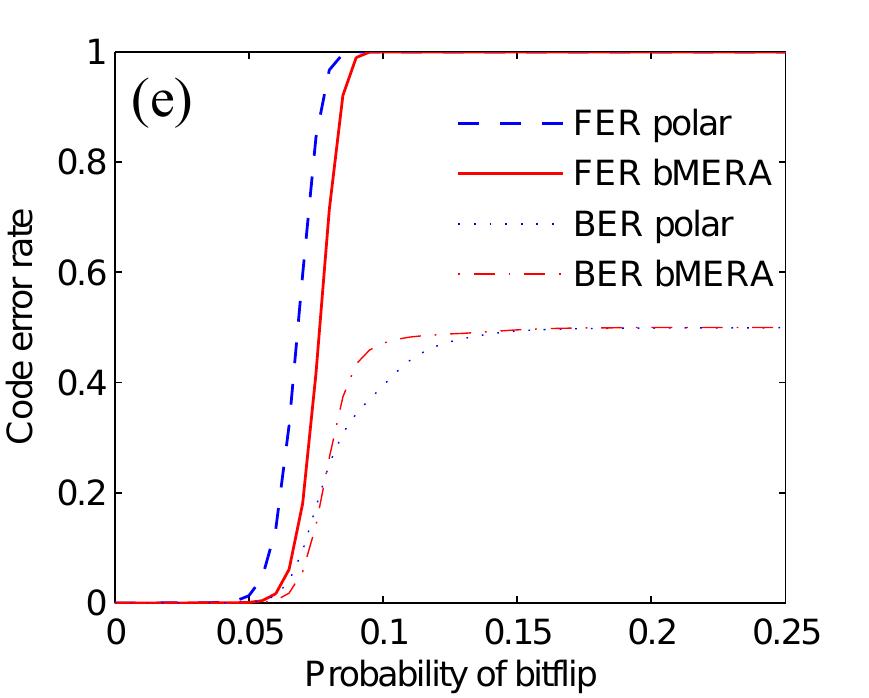}\includegraphics[width=0.5\doublecolumnwidth]{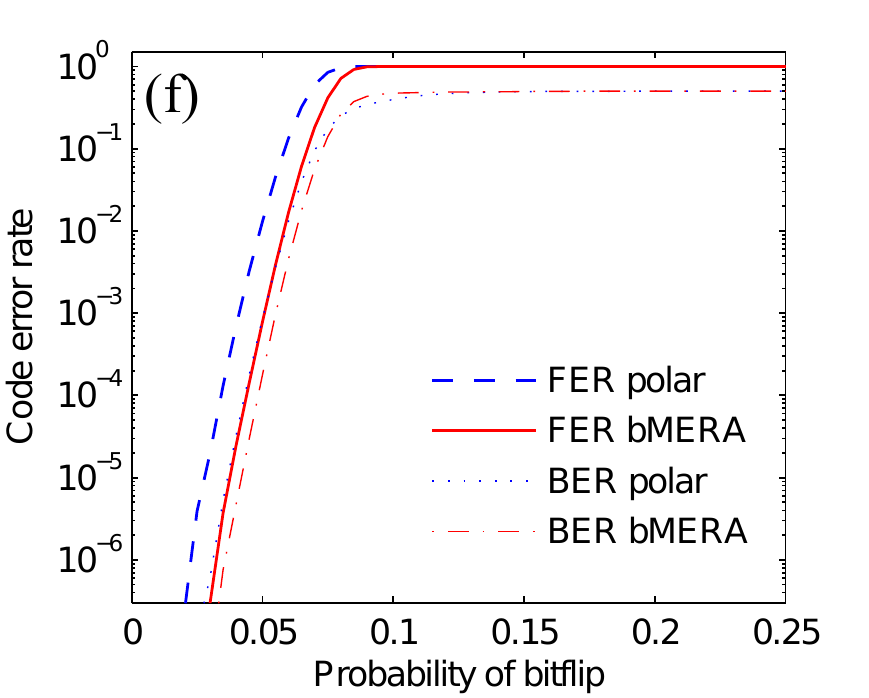}
\caption{Comparison of the performance of rate 1/2 Polar and Convolutional Polar codes of various sizes for the bit-flip channel. The encoded message contains (a,b) 256 bits. (c,d) 1024 bits and (e,f) 8192 bits. The capacity with bit-flip probability approximately 0.11 corresponds to the code rate 1/2. \label{fig:classical_bitflip} }
\end{figure}

In \fig{classical_bitflip} we see similar behaviour for the bit-flip channel. In this case we observe even greater finite-size effects, with the observed waterfall regions quite a bit below the expected threshold at a bit-flip rate of approximately 0.11. The Convolutional Polar code performs better in all cases, with a higher tolerance for error, a sharper transition between good and bad performance, and better scaling in the low error-rate region.

\begin{figure}[t]
\centering
\includegraphics[width=0.5\doublecolumnwidth]{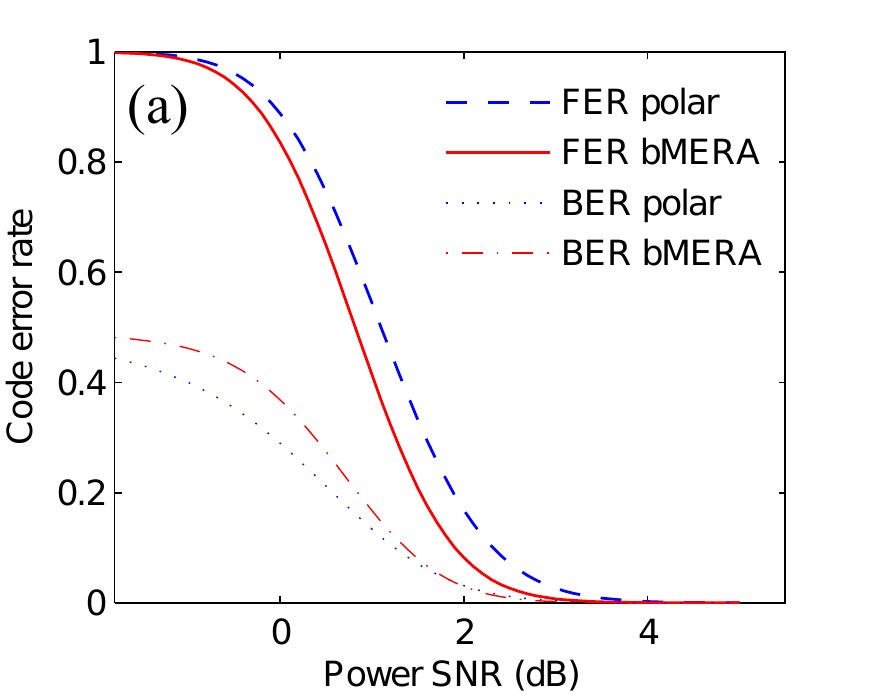}\includegraphics[width=0.5\doublecolumnwidth]{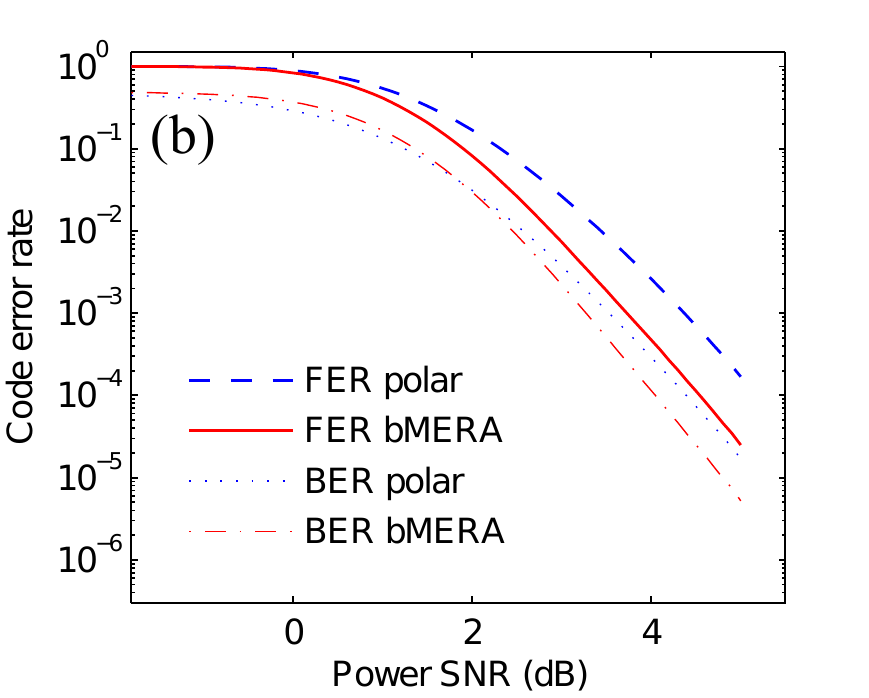}
\includegraphics[width=0.5\doublecolumnwidth]{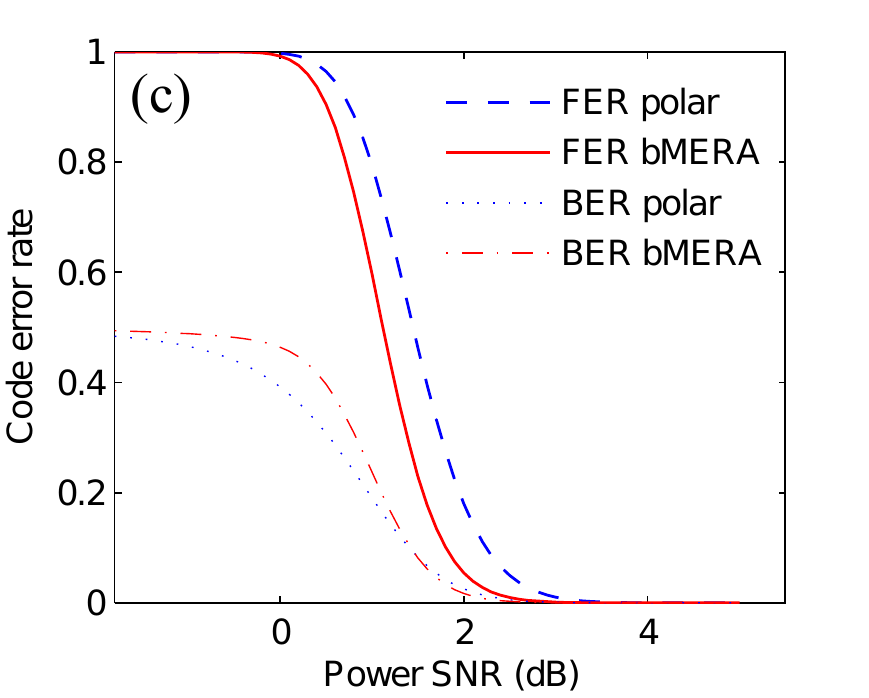}\includegraphics[width=0.5\doublecolumnwidth]{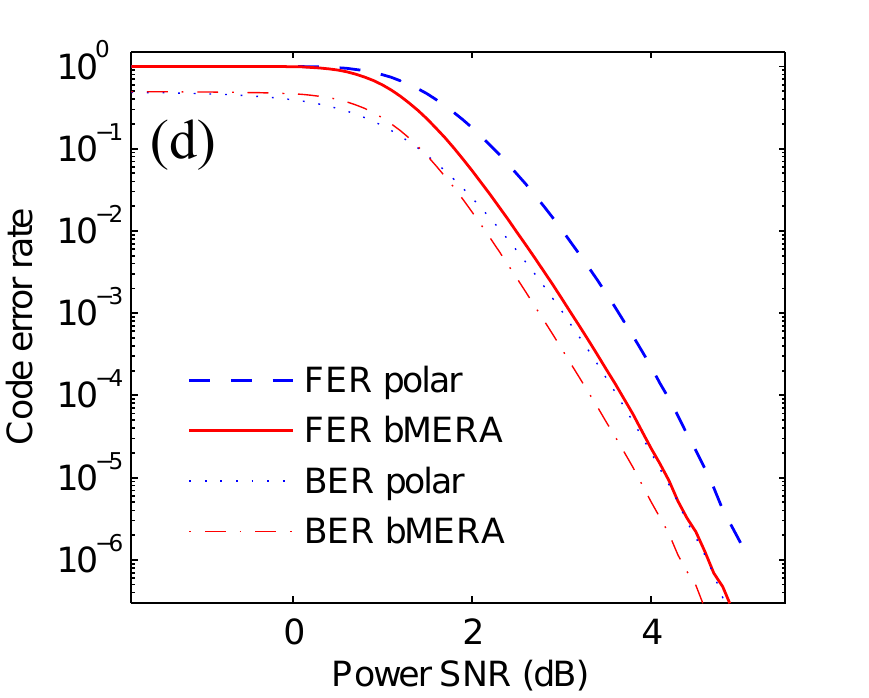}
\includegraphics[width=0.5\doublecolumnwidth]{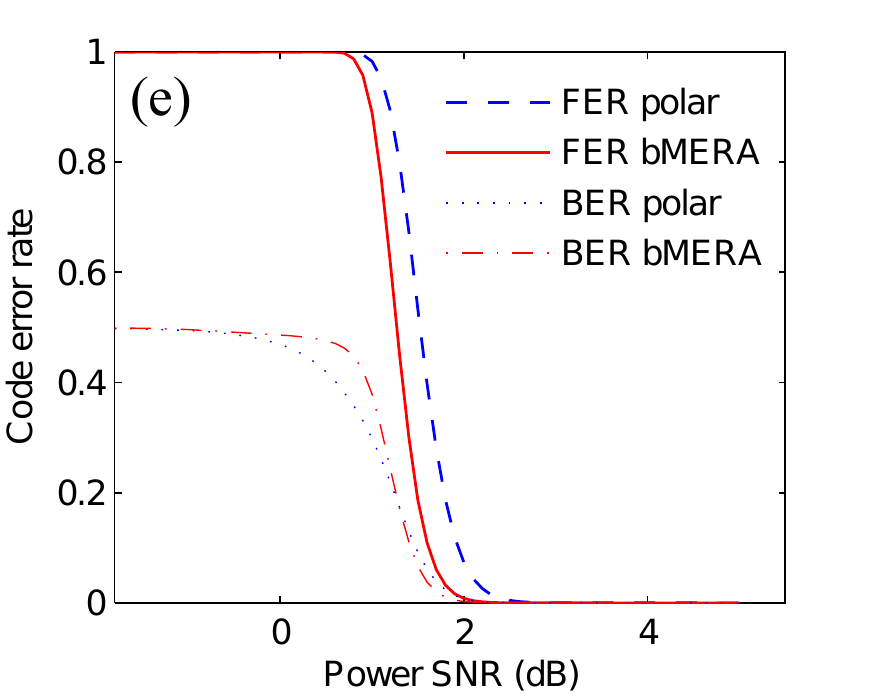}\includegraphics[width=0.5\doublecolumnwidth]{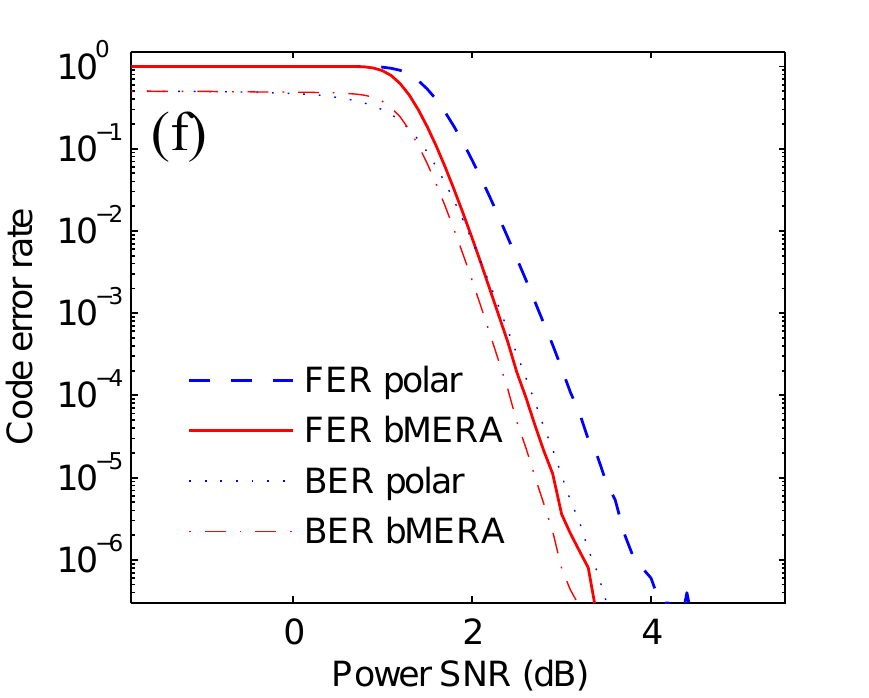}
\caption{Comparison of the performance of rate 1/2 Polar and Convolutional Polar codes of various sizes for the additive Gaussian white noise channel. The encoded message contains (a,b) 256 bits. (c,d) 1024 bits and (e,f) 8192 bits. \label{fig:classical_agwn} }
\end{figure}

Finally, we investigated performance under the more realistic additive Gaussian white noise channel in \fig{classical_agwn}. Once again we observe similar behaviour: the Convolutional Polar code has better error performance than the Polar code, including tolerance for larger noise rates and better scaling in the low noise region.

Based on these results, we can conclude that the Convolutional Polar code is a significant improvement to the Polar code when it comes to error correction capabilities with finite, relatively small block size. Also, the numerical cost is not changed significantly, with the same scaling and, anecdotally, approximately twice the computation effort to decode.

\section{Conclusion and Discussion}
\label{sec:discussion}

Using a family of tensor network recently introduced in the setting of quantum many-body physics, we presented in \cite{FP14}  a new family of error-correcting codes that generalize Polar codes in a  natural way. Recasting the decoding problem as a tensor network contraction, we have demonstrated that Arikan's sequential decoder can be realized with Convolutional Polar codes with log-linear complexity,  requiring roughly twice the computational effort of Polar codes sequential decoding. We have proved that Convolutional Polar codes achieve the capacity of any binary-input memoryless symmetric channel, and, moreover, that its error probability is provably  better than that of Polar codes.

For finite block size, our numerical simulations show that this new code outperforms Polar codes in several ways, including stronger channel polarization and enhanced error-correcting performances.
On the other hand, there clearly is more room for improvement, so that finite-size performance is closer to capacity. For instance, more complex schemes for channel selection may be possible. We have performed an additional analysis of the maximum likelihood decoder for smaller Polar and Convolutional Polar codes under the erasure channel and our results indicate performance significantly closer to capacity than observed in \fig{classical_erasure}. This difference arises because at all stages of decoding, every syndrome measurement is available to be used, unlike the successive cancellation decoder which only has access to previous bits. We speculate that, for small block sizes, the main advantage of Convolutional Polar code compared to the Polar code is that the syndrome bits are more tightly clustered to one side and the data channels on the other --- thus increasing the information available to the successive cancellation decoder for the earlier data bits. 

The previous observation is also tightly related to the \textit{universality} of the codes. We know that  Polar codes can be extended by additional constructions to become universal (in the sense, that they achieve the capacity of a compound channel)~\cite{SW13, HU13}. The main approaches either first polarize the channels and then combine them to make the code universal \cite{HU13} or first sort the channels by combining those with different mutual information by CNOT gates and then polarize~\cite{SW13}. Already in \cite{SW13} is mentioned that one would also expect an intermediate scheme to be possible. 
In this spirit Convolutional Polar codes can be seen as alternating between the sorting and the polarization steps from \cite{SW13} and indeed the channels appear to be more ordered. It would be interesting to further explore universality for Convolutional Polar codes or BMERA codes in general. 

The connection between tensor networks and coding opens the door to many other encoding schemes. The specific Convolutional Polar code studied here is only an example within the branching MERA family, many different codes can be obtained by varying the elementary gates in the network, including non-linear gates, and increasing the number of bits in elementary gates (i.e. increasing the ``bond dimension" in the TN language).  In particular, we saw that the increased exponent in \eq{W+} is responsible for the improved error exponent: further increases of this exponent are possible within the Convolutional Polar code family, albeit at an increased decoding complexity. Other tensor networks could also be considered along with their heuristic contraction schemes, e.g. \cite{VC04a,GLW08a}. In a similar vein, other decoders including belief propagation~\cite{Eslami2010} and list decoding~\cite{Tal2011} could also enhance the error-correction performances. 

A further interesting question is whether the results in this work can be extended to the realm of quantum information theory. 
Polar Codes have been shown to be capacity achieving also for classical-quantum channel \cite{WG13} and also the block error rate was shown to be identical to classical Polar codes \cite{WG13, H14}. The first question would be whether the results in this work can be extended to polarization for classical-quantum channels. A natural direction would be to follow the approach of this work, but no analog of the above used \textit{Mrs. Gerbers Lemma} with quantum side information is known up to now. 
Lastly, fully quantum versions of these codes can also be defined and similarly outperform quantum Polar codes \cite{FP142}. 
Nevertheless many open problems are left to fully compare the two codes. For Polar Codes we know many things, for example about their scaling behavior~\cite{GX15} and how to extended them to many coding (multi-user) settings in classical \cite{A12, STY13} and quantum information theory~\cite{HMW14,CM15}. 
All of these achievements naturally pose interesting questions for BMERA codes. 

\section*{Acknowledgments}
The authors would like to Jean-Pierre Tillich for useful discussions and Benjamin Bourassa for producing the data of \fig{error_exponent}. AJF would like to thank TOQATA (Spanish grant PHY008-00784), the EU IP SIQS, and the MPI-ICFO collaboration for supporting this research. CH acknowledges support by the Spanish MINECO, project FIS2013-40627-P and FPI Grant No. BES-2014-068888, as well as by the Generalitat de Catalunya, CIRIT project no. 2014 SGR 966. This work was funded by  NSERC (Natural Sciences and Engineering Research Council of Canada), CIfAR (The Canadian Institute for Advanced Research), and FRQNT (Fonds de recherche du Qu�bec � Nature et technologies) through INTRIQ. Computational resources were provided by Compute Canada and Calcul Quebec.

\appendix

We now elaborate on the concept of states of knowledge used for the erasure channel. For a single bit $x_1$ under the erasure channel, there are only two states of knowledge: we either know the value of $x_1$, or we do not. For two bits $x_1, x_2$ (under the erasure channel followed by a linear circuit) we know either both values $\{x_1, x_2\}$, a single value $\{x_1\}$ or $\{x_2\}$, their sum $\{x_1 + x_2\}$ or nothing --- so there are 5 possible states of knowledge each with its own probability of occurring. The application of a linear circuit (such as CNOT) permute the single-bit knowledge states $x_1$, $x_2$ and $x_1+x_2$, and their associated probabilities. In general, these states of knowledge correspond to the fact that erasure channels quantize trivially. 

The Convolutional Polar code under successive cancellation decoding has a tensor network with treewidth 3, and so we will need to deal with states of knowledge over 3 bits. For convenience we enumerate all 15 states below.
\begin{gather}
   s_1 = \emptyset \nonumber \\
   s_2 = \{x_1\} \nonumber \\
   s_3 = \{x_2\} \nonumber \\
   s_4 = \{x_3\} \nonumber \\
   s_5 = \{x_1 + x_2\} \nonumber \\
   s_6 = \{x_1 + x_3\} \nonumber \\
   s_7 = \{x_2 + x_3\} \nonumber \\
   s_8 = \{x_1 + x_2 + x_3\} \nonumber \\
   s_9 = \{ x_1, x_2 \} \nonumber \\
   s_{10} = \{ x_1, x_3 \} \nonumber \\
   s_{11} = \{ x_2, x_3 \} \nonumber \\
   s_{12} = \{ x_1, x_2 + x_3 \} \nonumber \\
   s_{13} = \{ x_2, x_1 + x_3 \} \nonumber \\
   s_{14} = \{ x_3, x_1 + x_2 \} \nonumber \\
   s_{15} = \{ x_1 + x_2, x_2 + x_3 \} \nonumber \\
   s_{16} = \{ x_1, x_2, x_3 \} 
\end{gather} 

We can determine how the probabilities $p_i$ associated with states $s_i$ transform under each layer of the Convolutional Polar code in the bulk, that is the transformation in Fig.~\ref{fig:contractions}~(b). Analogous to the Polar code, each layer of the Convolutional Polar  code combines two identical channels --- however in this case these are 3-bit channels. There are actually two distinct transformations different transformations given by the location of the 3 logical bits we are trying to decode. These are illustrated on the left and right of  Fig.~\ref{fig:contractions}~(b). These transformations are:

Figure \ref{fig:contractions}~(b) left:
\begin{subequations}
\begin{gather}
p^{\prime}_{1} = p_{1} p_{1} + p_{1} p_{2} + p_{1} p_{5} + p_{1} p_{6} + p_{1} p_{8} + p_{2} p_{1} + p_{5} p_{1} + p_{6} p_{1}\nonumber \\
 \hphantom{bla} + p_{8} p_{1}\\
p^{\prime}_{2} = p_{2} p_{2} + p_{5} p_{8} + p_{6} p_{6} + p_{8} p_{5}\\
p^{\prime}_{3} = p_{2} p_{8} + p_{5} p_{2} + p_{6} p_{5} + p_{8} p_{6}\\
p^{\prime}_{4} = p_{1} p_{4} + p_{1} p_{10} + p_{1} p_{14} + p_{2} p_{4} + p_{4} p_{1} + p_{4} p_{2} + p_{4} p_{4} + p_{4} p_{5}\nonumber \\
 \hphantom{bla} + p_{4} p_{6} + p_{4} p_{8} + p_{4} p_{10} + p_{4} p_{14} + p_{5} p_{4} + p_{6} p_{4} + p_{8} p_{4} + p_{10} p_{1}\nonumber \\
 \hphantom{bla} + p_{10} p_{4} + p_{14} p_{1} + p_{14} p_{4}\\
p^{\prime}_{5} = p_{1} p_{7} + p_{1} p_{12} + p_{1} p_{15} + p_{2} p_{7} + p_{3} p_{1} + p_{3} p_{2} + p_{3} p_{5} + p_{3} p_{6}\nonumber \\
 \hphantom{bla} + p_{3} p_{7} + p_{3} p_{8} + p_{3} p_{12} + p_{3} p_{15} + p_{5} p_{7} + p_{6} p_{7} + p_{8} p_{7} + p_{9} p_{1}\nonumber \\
 \hphantom{bla} + p_{9} p_{7} + p_{13} p_{1} + p_{13} p_{7}\\
p^{\prime}_{6} = p_{2} p_{6} + p_{5} p_{5} + p_{6} p_{2} + p_{8} p_{8}\\
p^{\prime}_{7} = p_{2} p_{5} + p_{5} p_{6} + p_{6} p_{8} + p_{8} p_{2}\\
p^{\prime}_{8} = p_{1} p_{3} + p_{1} p_{9} + p_{1} p_{13} + p_{2} p_{3} + p_{5} p_{3} + p_{6} p_{3} + p_{7} p_{1} + p_{7} p_{2}\nonumber \\
 \hphantom{bla} + p_{7} p_{3} + p_{7} p_{5} + p_{7} p_{6} + p_{7} p_{8} + p_{7} p_{9} + p_{7} p_{13} + p_{8} p_{3} + p_{12} p_{1}\nonumber \\
 \hphantom{bla} + p_{12} p_{3} + p_{15} p_{1} + p_{15} p_{3}\\
p^{\prime}_{9} = p_{2} p_{12} + p_{5} p_{12} + p_{6} p_{15} + p_{8} p_{15} + p_{9} p_{2} + p_{9} p_{8} + p_{9} p_{12} + p_{13} p_{5}\nonumber \\
 \hphantom{bla} + p_{13} p_{6} + p_{13} p_{15}\\
p^{\prime}_{10} = p_{2} p_{10} + p_{5} p_{14} + p_{6} p_{10} + p_{8} p_{14} + p_{10} p_{2} + p_{10} p_{6} + p_{10} p_{10} + p_{14} p_{5}\nonumber \\
 \hphantom{bla} + p_{14} p_{8} + p_{14} p_{14}\\
p^{\prime}_{11} = p_{2} p_{14} + p_{5} p_{10} + p_{6} p_{14} + p_{8} p_{10} + p_{10} p_{5} + p_{10} p_{8} + p_{10} p_{14} + p_{14} p_{2}\nonumber \\
 \hphantom{bla} + p_{14} p_{6} + p_{14} p_{10}\\
p^{\prime}_{12} = p_{2} p_{9} + p_{5} p_{13} + p_{6} p_{13} + p_{8} p_{9} + p_{12} p_{2} + p_{12} p_{5} + p_{12} p_{9} + p_{15} p_{6}\nonumber \\
 \hphantom{bla} + p_{15} p_{8} + p_{15} p_{13}\\
p^{\prime}_{13} = p_{2} p_{13} + p_{5} p_{9} + p_{6} p_{9} + p_{8} p_{13} + p_{12} p_{6} + p_{12} p_{8} + p_{12} p_{13} + p_{15} p_{2}\nonumber \\
 \hphantom{bla} + p_{15} p_{5} + p_{15} p_{9}\\
p^{\prime}_{14} = p_{1} p_{11} + p_{1} p_{16} + p_{2} p_{11} + p_{3} p_{3} + p_{3} p_{4} + p_{3} p_{9} + p_{3} p_{10} + p_{3} p_{11}\nonumber \\
 \hphantom{bla} + p_{3} p_{13} + p_{3} p_{14} + p_{3} p_{16} + p_{4} p_{3} + p_{4} p_{7} + p_{4} p_{9} + p_{4} p_{11} + p_{4} p_{12}\nonumber \\
 \hphantom{bla} + p_{4} p_{13} + p_{4} p_{15} + p_{4} p_{16} + p_{5} p_{11} + p_{6} p_{11} + p_{7} p_{4} + p_{7} p_{7} + p_{7} p_{10}\nonumber \\
 \hphantom{bla} + p_{7} p_{11} + p_{7} p_{12} + p_{7} p_{14} + p_{7} p_{15} + p_{7} p_{16} + p_{8} p_{11} + p_{9} p_{3} + p_{9} p_{4}\nonumber \\
 \hphantom{bla} + p_{9} p_{11} + p_{10} p_{3} + p_{10} p_{7} + p_{10} p_{11} + p_{11} p_{1} + p_{11} p_{2} + p_{11} p_{3} + p_{11} p_{4}\nonumber \\
 \hphantom{bla} + p_{11} p_{5} + p_{11} p_{6} + p_{11} p_{7} + p_{11} p_{8} + p_{11} p_{9} + p_{11} p_{10} + p_{11} p_{11} + p_{11} p_{12}\nonumber \\
 \hphantom{bla} + p_{11} p_{13} + p_{11} p_{14} + p_{11} p_{15} + p_{11} p_{16} + p_{12} p_{4} + p_{12} p_{7} + p_{12} p_{11} + p_{13} p_{3}\nonumber \\
 \hphantom{bla} + p_{13} p_{4} + p_{13} p_{11} + p_{14} p_{3} + p_{14} p_{7} + p_{14} p_{11} + p_{15} p_{4} + p_{15} p_{7} + p_{15} p_{11}\nonumber \\
 \hphantom{bla} + p_{16} p_{1} + p_{16} p_{3} + p_{16} p_{4} + p_{16} p_{7} + p_{16} p_{11}\\
p^{\prime}_{15} = p_{2} p_{15} + p_{5} p_{15} + p_{6} p_{12} + p_{8} p_{12} + p_{9} p_{5} + p_{9} p_{6} + p_{9} p_{15} + p_{13} p_{2}\nonumber \\
 \hphantom{bla} + p_{13} p_{8} + p_{13} p_{12}\\
p^{\prime}_{16} = p_{2} p_{16} + p_{5} p_{16} + p_{6} p_{16} + p_{8} p_{16} + p_{9} p_{9} + p_{9} p_{10} + p_{9} p_{13} + p_{9} p_{14}\nonumber \\
 \hphantom{bla} + p_{9} p_{16} + p_{10} p_{9} + p_{10} p_{12} + p_{10} p_{13} + p_{10} p_{15} + p_{10} p_{16} + p_{12} p_{10} + p_{12} p_{12}\nonumber \\
 \hphantom{bla} + p_{12} p_{14} + p_{12} p_{15} + p_{12} p_{16} + p_{13} p_{9} + p_{13} p_{10} + p_{13} p_{13} + p_{13} p_{14} + p_{13} p_{16}\nonumber \\
 \hphantom{bla} + p_{14} p_{9} + p_{14} p_{12} + p_{14} p_{13} + p_{14} p_{15} + p_{14} p_{16} + p_{15} p_{10} + p_{15} p_{12} + p_{15} p_{14}\nonumber \\
 \hphantom{bla} + p_{15} p_{15} + p_{15} p_{16} + p_{16} p_{2} + p_{16} p_{5} + p_{16} p_{6} + p_{16} p_{8} + p_{16} p_{9} + p_{16} p_{10}\nonumber \\
 \hphantom{bla} + p_{16} p_{12} + p_{16} p_{13} + p_{16} p_{14} + p_{16} p_{15} + p_{16} p_{16}
\end{gather}
\end{subequations}

\newpage

Figure \ref{fig:contractions}~(b) right:
\begin{subequations}
\begin{gather}
p^{\prime}_{1} = p_{1} p_{1} + p_{1} p_{2} + p_{1} p_{3} + p_{1} p_{5} + p_{1} p_{6} + p_{1} p_{7} + p_{1} p_{8} + p_{1} p_{9}\nonumber \\
 \hphantom{bla} + p_{1} p_{12} + p_{1} p_{13} + p_{1} p_{15} + p_{2} p_{1} + p_{2} p_{2} + p_{2} p_{3} + p_{2} p_{6} + p_{2} p_{7}\nonumber \\
 \hphantom{bla} + p_{3} p_{1} + p_{3} p_{2} + p_{3} p_{5} + p_{3} p_{6} + p_{3} p_{8} + p_{5} p_{1} + p_{5} p_{3} + p_{5} p_{5}\nonumber \\
 \hphantom{bla} + p_{5} p_{7} + p_{5} p_{8} + p_{6} p_{1} + p_{6} p_{2} + p_{6} p_{3} + p_{6} p_{6} + p_{6} p_{7} + p_{7} p_{1}\nonumber \\
 \hphantom{bla} + p_{7} p_{2} + p_{7} p_{5} + p_{7} p_{6} + p_{7} p_{8} + p_{8} p_{1} + p_{8} p_{3} + p_{8} p_{5} + p_{8} p_{7}\nonumber \\
 \hphantom{bla} + p_{8} p_{8} + p_{9} p_{1} + p_{12} p_{1} + p_{13} p_{1} + p_{15} p_{1}\\
p^{\prime}_{2} = p_{5} p_{2} + p_{5} p_{9} + p_{5} p_{12} + p_{8} p_{6} + p_{8} p_{13} + p_{8} p_{15} + p_{9} p_{2} + p_{12} p_{6}\nonumber \\
 \hphantom{bla} + p_{13} p_{6} + p_{15} p_{2}\\
p^{\prime}_{3} = p_{3} p_{3} + p_{3} p_{9} + p_{3} p_{13} + p_{7} p_{7} + p_{7} p_{12} + p_{7} p_{15} + p_{9} p_{3} + p_{12} p_{7}\nonumber \\
 \hphantom{bla} + p_{13} p_{3} + p_{15} p_{7}\\
p^{\prime}_{4} = p_{3} p_{7} + p_{3} p_{12} + p_{3} p_{15} + p_{7} p_{3} + p_{7} p_{9} + p_{7} p_{13} + p_{9} p_{7} + p_{12} p_{3}\nonumber \\
 \hphantom{bla} + p_{13} p_{7} + p_{15} p_{3}\\
p^{\prime}_{5} = p_{2} p_{5} + p_{2} p_{9} + p_{2} p_{15} + p_{6} p_{8} + p_{6} p_{12} + p_{6} p_{13} + p_{9} p_{5} + p_{12} p_{5}\nonumber \\
 \hphantom{bla} + p_{13} p_{8} + p_{15} p_{8}\\
p^{\prime}_{6} = p_{2} p_{8} + p_{2} p_{12} + p_{2} p_{13} + p_{6} p_{5} + p_{6} p_{9} + p_{6} p_{15} + p_{9} p_{8} + p_{12} p_{8}\nonumber \\
 \hphantom{bla} + p_{13} p_{5} + p_{15} p_{5}\\
p^{\prime}_{7} = p_{1} p_{4} + p_{1} p_{10} + p_{1} p_{11} + p_{1} p_{14} + p_{1} p_{16} + p_{2} p_{4} + p_{2} p_{10} + p_{2} p_{11}\nonumber \\
 \hphantom{bla} + p_{3} p_{4} + p_{3} p_{10} + p_{3} p_{14} + p_{4} p_{1} + p_{4} p_{2} + p_{4} p_{3} + p_{4} p_{4} + p_{4} p_{5}\nonumber \\
 \hphantom{bla} + p_{4} p_{6} + p_{4} p_{7} + p_{4} p_{8} + p_{4} p_{9} + p_{4} p_{10} + p_{4} p_{11} + p_{4} p_{12} + p_{4} p_{13}\nonumber \\
 \hphantom{bla} + p_{4} p_{14} + p_{4} p_{15} + p_{4} p_{16} + p_{5} p_{4} + p_{5} p_{11} + p_{5} p_{14} + p_{6} p_{4} + p_{6} p_{10}\nonumber \\
 \hphantom{bla} + p_{6} p_{11} + p_{7} p_{4} + p_{7} p_{10} + p_{7} p_{14} + p_{8} p_{4} + p_{8} p_{11} + p_{8} p_{14} + p_{9} p_{4}\nonumber \\
 \hphantom{bla} + p_{10} p_{1} + p_{10} p_{2} + p_{10} p_{3} + p_{10} p_{4} + p_{10} p_{6} + p_{10} p_{7} + p_{10} p_{10} + p_{10} p_{11}\nonumber \\
 \hphantom{bla} + p_{11} p_{1} + p_{11} p_{2} + p_{11} p_{4} + p_{11} p_{5} + p_{11} p_{6} + p_{11} p_{8} + p_{11} p_{10} + p_{11} p_{14}\nonumber \\
 \hphantom{bla} + p_{12} p_{4} + p_{13} p_{4} + p_{14} p_{1} + p_{14} p_{3} + p_{14} p_{4} + p_{14} p_{5} + p_{14} p_{7} + p_{14} p_{8}\nonumber \\
 \hphantom{bla} + p_{14} p_{11} + p_{14} p_{14} + p_{15} p_{4} + p_{16} p_{1} + p_{16} p_{4}\\
p^{\prime}_{8} = p_{5} p_{6} + p_{5} p_{13} + p_{5} p_{15} + p_{8} p_{2} + p_{8} p_{9} + p_{8} p_{12} + p_{9} p_{6} + p_{12} p_{2}\nonumber \\
 \hphantom{bla} + p_{13} p_{2} + p_{15} p_{6}\\
p^{\prime}_{9} = p_{9} p_{9} + p_{12} p_{15} + p_{13} p_{13} + p_{15} p_{12}\\
p^{\prime}_{10} = p_{9} p_{12} + p_{12} p_{13} + p_{13} p_{15} + p_{15} p_{9}\\
p^{\prime}_{11} = p_{3} p_{11} + p_{3} p_{16} + p_{7} p_{11} + p_{7} p_{16} + p_{9} p_{11} + p_{11} p_{3} + p_{11} p_{7} + p_{11} p_{9}\nonumber \\
 \hphantom{bla} + p_{11} p_{11} + p_{11} p_{12} + p_{11} p_{13} + p_{11} p_{15} + p_{11} p_{16} + p_{12} p_{11} + p_{13} p_{11} + p_{15} p_{11}\nonumber \\
 \hphantom{bla} + p_{16} p_{3} + p_{16} p_{7} + p_{16} p_{11}\\
p^{\prime}_{12} = p_{5} p_{10} + p_{5} p_{16} + p_{8} p_{10} + p_{8} p_{16} + p_{9} p_{10} + p_{12} p_{10} + p_{13} p_{10} + p_{14} p_{2}\nonumber \\
 \hphantom{bla} + p_{14} p_{6} + p_{14} p_{9} + p_{14} p_{10} + p_{14} p_{12} + p_{14} p_{13} + p_{14} p_{15} + p_{14} p_{16} + p_{15} p_{10}\nonumber \\
 \hphantom{bla} + p_{16} p_{2} + p_{16} p_{6} + p_{16} p_{10}\\
p^{\prime}_{13} = p_{9} p_{13} + p_{12} p_{12} + p_{13} p_{9} + p_{15} p_{15}\\
p^{\prime}_{14} = p_{9} p_{15} + p_{12} p_{9} + p_{13} p_{12} + p_{15} p_{13}\\
p^{\prime}_{15} = p_{2} p_{14} + p_{2} p_{16} + p_{6} p_{14} + p_{6} p_{16} + p_{9} p_{14} + p_{10} p_{5} + p_{10} p_{8} + p_{10} p_{9}\nonumber \\
 \hphantom{bla} + p_{10} p_{12} + p_{10} p_{13} + p_{10} p_{14} + p_{10} p_{15} + p_{10} p_{16} + p_{12} p_{14} + p_{13} p_{14} + p_{15} p_{14}\nonumber \\
 \hphantom{bla} + p_{16} p_{5} + p_{16} p_{8} + p_{16} p_{14}\\
p^{\prime}_{16} = p_{9} p_{16} + p_{12} p_{16} + p_{13} p_{16} + p_{15} p_{16} + p_{16} p_{9} + p_{16} p_{12} + p_{16} p_{13} + p_{16} p_{15}\nonumber \\
 \hphantom{bla} + p_{16} p_{16}
\end{gather}
\end{subequations}

Our evaluation of the erasure probability of \sec{erasure} simply iterates these transformations following the contraction schedule of the tensor network associated to the code.



\end{document}